\documentclass[%
 aip,
 amsmath,amssymb,
reprint,
]{revtex4-1}

\usepackage{graphicx}
\usepackage{dcolumn}
\usepackage{bm}

\usepackage{algpseudocode,algorithm,algorithmicx}
\usepackage{hyperref}

\usepackage[utf8]{inputenc}
\usepackage[T1]{fontenc}
\usepackage{mathptmx}
\usepackage{xcolor}
\begin{document}


\title[]{Constrained non-negative matrix factorization enabling real-time insights of \emph{in situ} and high-throughput experiments}

\author{Phillip M. Maffettone}
\affiliation{National Synchrotron Light Source II, Brookhaven National Laboratory, Upton, NY 11973, USA}
\author{Aidan C. Daly}%
\affiliation{ 
Center for Computational Biology, Flatiron Institute, U.S.A
}%

\author{Daniel Olds}
 \email{dolds@bnl.gov}
\affiliation{National Synchrotron Light Source II, Brookhaven National Laboratory, Upton, NY 11973, USA}%

\date{\today}

\begin{abstract}
%
%
Non-negative Matrix Factorization (NMF) methods offer an appealing unsupervised learning method for real-time analysis of streaming spectral data in time-sensitive data collection, such as \textit{in situ} characterization of materials.  However, canonical NMF methods are optimized to reconstruct a full dataset as closely as possible, with no underlying requirement that the reconstruction produces components or weights representative of the true physical processes.  
In this work, we demonstrate how constraining NMF weights or components, provided as known or assumed priors, can provide significant improvement in revealing true underlying phenomena.  
We present a PyTorch based method for efficiently applying constrained NMF and demonstrate this on several synthetic examples.  
When applied to streaming experimentally measured spectral data, an expert researcher-in-the-loop can provide and dynamically adjust the constraints. 
This set of interactive priors to the NMF model can, for example, contain known or identified independent components, as well as functional expectations about the mixing of components.
 We demonstrate this application on measured X-ray diffraction and pair distribution function data from \textit{in situ} beamline experiments.
Details of the method are described, and general guidance provided to employ constrained NMF in extraction of critical information and insights during \textit{in situ} and high-throughput experiments.
\end{abstract}

\maketitle


\section{\label{sec:intro}Introduction}
High-throughput\cite{Kusne_2014, Gomez-Bombarelli_2016, Greenaway_2018, Langner_2020, Ludwig_2019, Decker_2017} and \emph{in situ}\cite{Yeung_2016, MacLeod_2020} analyses have become ubiquitous in modern materials discovery. 
Automation and engineering has increased capabilities for data streaming and \emph{in situ} experimentation in many settings, including university laboratories,\cite{Li_2020} industrial research,\cite{Fleischer_2018} and  central facilities.\cite{Campbell_2020} 
This technological advancement opens up new opportunities for interrogating the dynamic processes of a system including phase transitions\cite{Olds_2017} and reactive processes.\cite{Yeung_2016}
Such opportunities bear the burden of an increased rate of data production, as well as a commensurate increase in the volume of data to process. 
For many experiments, these high-volume datasets require physically meaningful analysis to be performed in real-time. 
Diffraction experiments are a prototypical materials characterization technique that produce such data. 
X-ray diffraction (XRD) and pair distribution function (PDF) analysis are necessary for comprehending energy materials across batteries,\cite{Hua_2021, Shadike_2021} catalysis,\cite{Wang_2019, Loffler_2019} photovoltaics,\cite{Olds_2020} and multi-ferroics.\cite{Clarke_2021}
This challenge is most pressing at central facilities, which have the technological capacity to produce vast amounts of data across variable sample environments. 
\par

With the advances in accelerator technology achieved in recent decades, available photon flux at synchrotron X-ray lightsources is increasing at a rate even faster than Moore’s law.\cite{Olds_2020}
With new beamlines capable of utilizing these record-breaking levels of brightness and resolution, scientists can now perform many of the critical high-throughput and \emph{in situ} studies required to realize future energy technologies.  
Largely driven by these new capabilities, facilities are generating more data now than ever before. 
For example, at the National Synchrotron Light Source II (NSLS-II) produced over 1 PB of raw data in the last year, and that rate is expected to increase as the facility matures.\cite{Campbell_2020} 
Despite these huge data generation rates, however, approaches to data analysis have barely changed over time, with users often relying on methods and code that are decades old.
Consequently, data collected in seconds to minutes may take weeks or months of analysis to understand. 
Due to such limitations, conventional knowledge extraction is often divorced from the beamtime measurement process.  This lack of real-time feedback can force users into ‘flying blind’ during their limited time at the beamline, leading to missed opportunities and mistakes.  

This information bottleneck is most pronounced during diffraction experiments, where the atomic structure of a sample can be measured in fractions of a second, in both real (PDF) and reciprocal (XRD) space.
Historically, such data are modeled using iterative refinement procedures such as Rietveld refinements to extract the relevant physical insights such as crystallographic phase, lattice constants, atomic positions, site occupancies, and atomic displacement parameters.\cite{Rietveld_1967}  
As these refinements may take minutes to hours for a skilled expert to perform, they are daunting for technique novices and often do not scale well to large datastreams.\cite{Giacovazzo_2011}
While such full information extraction from the data can be performed \emph{a posteriori}, users distinctly benefit from the extraction of critical information at-the-beamline, such as if the sample has undergone a sought transition or phase transformation.
\par

Machine learning (ML) tools offer great promise for monitoring and analyzing large volumes of streaming data.\cite{Campbell_2020}
These tools have been demonstrated for classifying diffraction patterns,\cite{Iwasaki_2017, Xiong_2017, Long_2009, Oviedo_2019, Lee_2020, Ziletti_2018, Aguiar_2019, Chen_2020} or decomposing large datasets.\cite{Takeuchi_2005, Olds_2017, Stanev_2018, Bai_2017, Bai_2018, Suram_2017} 
Particularly, unsupervised learning is useful for offering rapid diagnostics and analysis without prior training.\cite{Hua_2021, Geddes_2019, Stanev_2018} 
Because unsupervised decomposition---or data reduction---algorithms do not require any prior data labeling or expected phases, they are exceptionally useful for interpreting datasets where there are potentially new or unexpected phases.
Two such algorithms, principle component analysis (PCA)\cite{Olds_2017} and non-neagative matrix fatorization (NMF),\cite{Geddes_2019, Li_2020} have been employed to segregate components from diffraction datasets. The algorithms aim to describe a dataset of $m$ experimental patterns (XRD or PDF) as a weighted linear composition of $k$ fundamental components.
Of these, NMF is particularly well suited to scattering data because its decomposition aligns with physics: the component signals are enforced to remain greater than or equal to zero, and are strictly additive to one another in the dataset reconstruction.
\par

While NMF has the advantages of retaining the additive superposition and non-negativity of XRD and PDF, it comes with some significant drawbacks. 
Firstly, NMF is incapable of interpreting translational variance. Since the NMF model uses only a linear combination of components, a given component cannot effectively reproduce data that is identical except for translation. Unfortunately this is a common occurrence for XRD and PDF experiments due to thermal expansion, changes in nanostructure, and lattice strain due to Vergad's law.\cite{Stanev_2018}
Efforts to overcome this translation have included discarding translationally redundant components, although this requires additional hyperparameters for detecting the redundancy that are not effective across different experimental challenges.\cite{Iwasaki_2017, Stanev_2018}
Furthermore, there is no unique set of components and weights that can satisfy the NP-hard optimization for the NMF decomposition. This means that the NMF result will vary based on random initialization, and is not guaranteed to converge to a local minimum for reconstruction error. 
Lastly, while non-negativity is a sensible constraint, it doesn't require the resulting components to correspond to interpretable patterns nor the weights to carry any meaning with respect to volume fraction.
It can be argued that there is a unique set of components in the analysis of diffraction data that is physically meaningful. Metropolis matrix factorization attempts to identify this unique set by constraining the constraining select components and finding the global optimum for the constrained decomposition.\cite{Geddes_2019, Hua_2021}
Although this approach is powerful, it remains too costly to run real-time during an measurement that is streaming data. 
\par

To successfully apply decomposition methods during a diffraction experiment, they must operate efficiently and resolve components and weights that are physically meaningful. 
Instead of simply finding the decomposition model that best fits the data, we need the \emph{most likely physical} model that best fits the data.
By increasing the likelihood that found components and weights are physically relevant to real chemical process occurring within the sample, it would present results in an immediately convincing representation to the researcher.  
For example, a sudden shift in weights between two components may suggest to a researcher that a second order phase transition has occurred during an experiment; however, if the researcher is presented with nonphysical components to inspect, they are far less likely to understand or trust the information even if it later proves true.  
Fortunately, the nature of these \emph{in situ} experiments typically follows a standard trajectory across a set of state variables, under which the sample begins readily known phase to the researcher that may undergo a transition at some future point into known or unknown phases.\cite{Li_2021}
Various combinations of known and unknown components or weights can be occur in a data series, such as knowledge of two components at the edges of a dataset with curiosity about a third, or no knowledge of the components, but expectations for their concentrations.\cite{Geddes_2019}
Thus, through use of constraints we can employ NMF in highly productive and insightful ways for diffraction experiments.
\par

In this contribution, we present our advances developing constrained NMF methods for use on \emph{in situ} studies at total scattering and powder diffraction beamlines. 
The algorithms are developed using the PyTorch\cite{Paszke2019} framework for speed and scalability.
We first demonstrate the capabilities of constrained NMF on several synthetic examples, in which the limitations of canonical NMF are explored. 
We then use constrained NMF to decompose temperature dependent XRD and PDF datasets of a ferroelectric perovskite and a molten salt, wherein there is no  \emph{a priori} knowledge of components in the latter.
Lastly, we provide best practices on how to apply these methods during \emph{in situ} experiments and interpret the results as they relate to commonly encountered scenarios in the analysis of energy materials.
These developments are extensible to applications in spectroscopy and microscopy, and can also be deployed at university and industrial laboratories.  
Within each of these results, constrained NMF behaves as a companion agent in concert with the researcher, leveraging the expertise of the scientist while providing real-time insights.

\section{\label{sec:results}Results and Discussion}
The goal of NMF is to find an $m \times k$ non-negative matrix $\mathbf{W}$ of weights and a $k \times n$ non-negative matrix $\mathbf{H}$ of components, to approximate a dataset $\mathbf{X} \approx \mathbf{WH}$, where $m$ is the number of patterns in a dataset, $n$ is the number of points in a pattern, and $k$ is the chosen number of components. The details of this optimization are presented in Section~\ref{sec:methods}. 
We augmented this approach to allow the user --- or an algorithm --- to initialize any column of $\mathbf{W}$ or row of $\mathbf{H}$, and to subsequently constrain those vectors from being updated during the optimization. 
It is worth noting that the algorithm is often presented with the contents of $\mathbf{W}$ and  $\mathbf{H}$ reversed; however, we find more intuitive to have $\mathbf{W}$ refer to weights. 
\par

\subsection{\label{sec:datasets} Synthetic and experimental datasets}

\begin{figure}
    \includegraphics{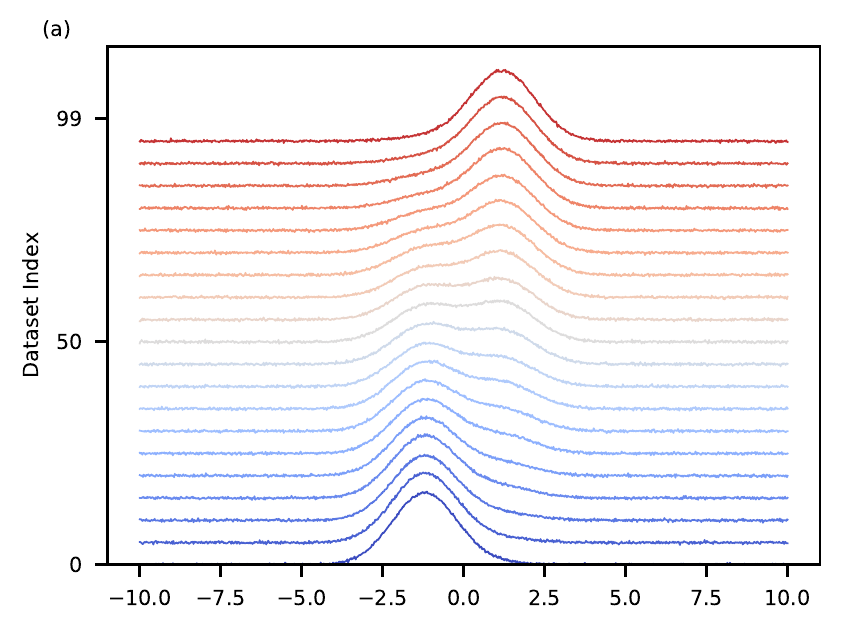}
    \includegraphics{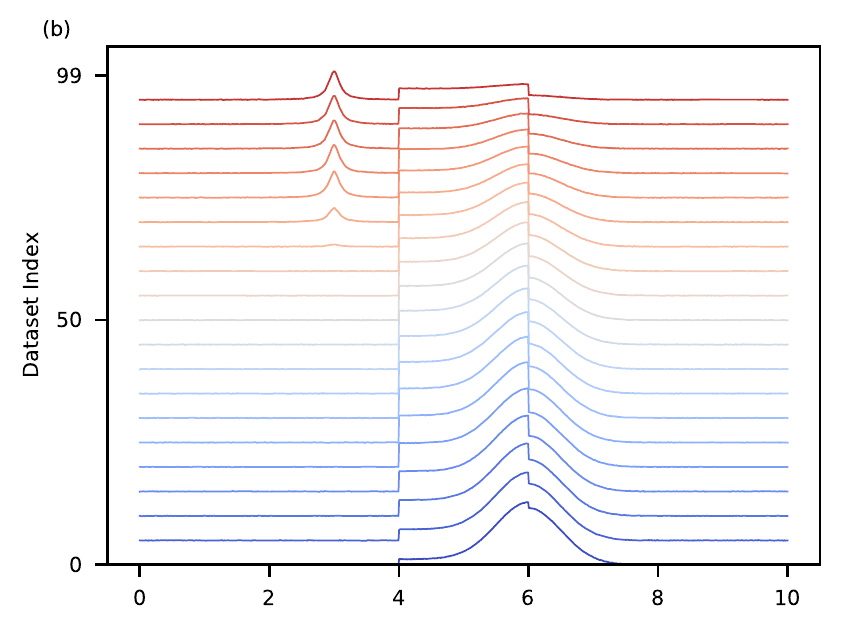}
    \caption{\label{fig:datasets}Example datasets used to demonstrate the capabilities of canonical and constrained NMF.
    (a) Two noisy Gaussian curves centered at 2.5 and -2.5 are mixed in linear combinations with coefficients that vary approximately, but not exactly, linearly across the dataset.
    (b) A Lorentzian centered at 3.0, a box function of width two centered at 5.0, and a Gaussian centered at 6.0 are mixed as linear combination with functionally varying weights across the dataset.}
\end{figure}

To test the capabilities of canonical and constrained NMF, we assembled two synthetic datasets (Fig.~\ref{fig:datasets}). The first approach highlights a common case where two unique, overlapping, components will mix linearly with coefficients that sum to one. This occurs in binary mixtures when the measured response function is a molar weighted sum of the pure response functions, as is the case for diffraction.  We represent this with two Gaussians mixed linearly, and include random noise in coefficients of mixing and the final functions.
Under these circumstances, it is likely to know the weights of the individual components \emph{a priori}, but not necessarily the components themselves. 
\par

The second synthetic dataset contains three unique functions that are mixed non-monotonically as the dataset progresses. 
The three functions are a non-overlapping Lorentzian, and a Gaussian and box function that overlap to create non-differentiable result.
This dataset explores a limiting and challenging use case for the underlying components, that can occur with a known mask or background augmentation.
The weights represent commonly encountered ways that physical and chemical processes --- and therefore components --- can evolve naturally. 
This family of challenges highlights the scenario wherein \emph{a priori} knowledge of one or all components exists, but not necessarily knowledge the weights. 
\par

\begin{figure}
    \includegraphics[]{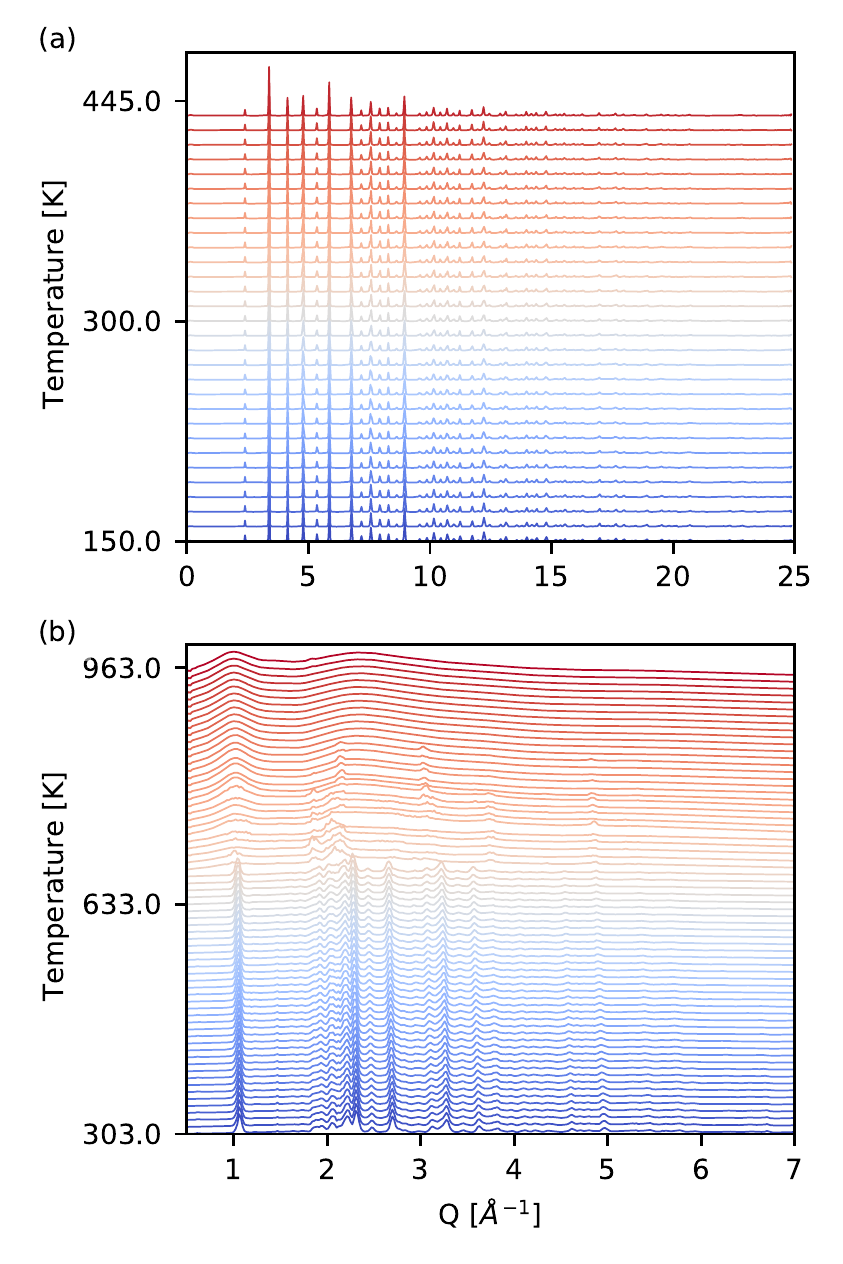}
    \caption{\label{fig:both_exp} Constrained NMF was used to analyze two variable temperature diffraction datasets. (a)  BaTiO$_3$ undergoes three phase transitions across the temperature range 150--450~\,K, that are imperceptible even to traditional refinement techniques. (b) NaCl:CrCl$_3$ diffraction was measured  a temperature range of 300--963\,K that included a melting transition and an anomolous high temperature solid phase (shown by the remaining sharp peaks above 650~\,K).}
\end{figure}

We next examined to two distinct experimental challenges, analyzing the exceedingly subtle temperature driven phase transitions of barium titanate, and understanding the structural evolution of a salt, NaCl:CrCl$_3$,  heated beyond it's melting temperature. 
BaTiO$_3$ is a well studied material that has been used to probe the mechanism of ferroelecticity because its phase transitions exist at relatively low temperature.\cite{Page_2010, Wegner_2020}
At high temperature, BaTiO$_3$ exists as a paraelectric cubic phase, where on average the Ti atoms are octahedrally coordinated by six O atoms. At room temperature the average displacement of the Ti elongates along [001], resulting in a ferroelectric tetragonal phase. Further cooling results in an orthorhombic phase with polarization along the [011]. At low temperature the rhombohedral ground state is reached, with all Ti atoms polarized along the [111].
This symmetry breaking results in subtle peak splitting, which is further obfuscated by thermal expansion driven peak shifting (Fig.~\ref{fig:both_exp}).
The differences between patterns is so subtle that classification across these temperature-induced phase transitions is not possible considering only Rietveld refinements without expert intervention.\cite{Page_2010}
\par

Contrary to changes in BaTiO$_3$, an crystalline--amorphous transition is abrupt and distinct.
We used constrained NMF to understand the behavior of NaCl:CrCl$_3$ across a temperature range of 300--963\,K, which included its melting transition (Fig.~\ref{fig:both_exp}). These molten salts are of interest for the development of nuclear reactors.\cite{Li_2021}
In this experiment there was a residual solid phase after the melting transition. Without any additional analysis, it remained unclear to the research team whether that solid was a distinct high-temperature phase or an experimental artefact. Together, these four datasets present the clear utility of constrained NMF for on-the-fly and \emph{in situ} analysis of diffraction data.

\subsection{\label{sec:toy_weights}Constraining weights with synthetic datasets}
With the first synthetic dataset (Fig.\ref{fig:datasets}a) we demonstrated the challenge of retaining physically relevant components from NMF when the weights are readily available. This is a common situation in diffraction studies of variable composition or phase mixing, when other experimental information (preparation conditions or spectroscopy) will indicate the concentration of individual phases without specifying the diffraction patterns of the phases. We considered a continuous mixture of two Gaussian functions, where noise is added to the composition and to the final function. From this we can see the power of canonical NMF for dimensionality reduction and the benefit offered by constraints. 
\par

\begin{figure}
    \includegraphics[]{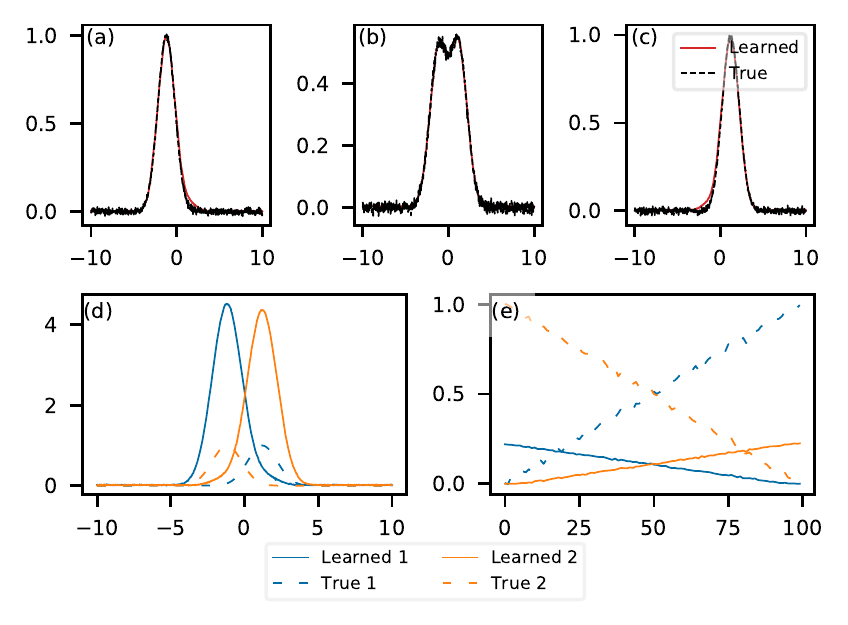}
    \caption{\label{fig:toy_gaussian} The reconstruction of the (a) first, (b) median, and (c) last pattern of a dataset of mixed Gaussian functions using canonical NMF. The full dataset is shown in Figure~\ref{fig:datasets}(a). 
    (d) Both the ordering and amplitude of the learned components fails to match the ground truth functions.
    (e) The learned weights compensate for this false magnitude and fit accurately to the non-linearity in the true weights. }
\end{figure}

In decomposing the Gaussian dataset (Fig~\ref{fig:datasets}(a)) using canonical NMF, the reconstruction loss as defined by squared error (SE) (eq.~\ref{eq:squared_error}) achieved a mean value (MSE) of 5.788 over the dataset. The common crystallographic metric of R-factor (eq.~\ref{eq:R}) was 0.066 for the reconstruction on average. 
The decomposition into two components created a model of the mixtures that fits the data well. Even though the learned components are similar to the true functions, their respective magnitudes are not obviously linked to the magnitudes of those functions (Fig~\ref{fig:toy_gaussian}d).
Furthermore, the weights must compensate for this increased predicted magnitude, and inevitably over-fit the noise in the dataset and  (Fig~\ref{fig:toy_gaussian}e). 
This is the expected behavior of NMF, but is naive to any prior knowledge or physical constraints we can impose on the dataset. 
\par

However, when we informed the NMF model with prior knowledge of weights, and constrained those weights, we constructed the most likely model that best fits the data. This resultant model successfully retains the physical significance on the weights and components. 
In this use case, the weights were limited to the coordinates of a simplex (that is, they sum to unity), and expected to vary linearly. 
By initializing the NMF with these exact weights, and constraining their update in the optimization algorithm, we created a decomposition with a MSE of 6.903 and mean R-factor 0.0722. 
By design these metrics are slightly worse than those for the canonical NMF because we do not allow the decomposition to over-fit the noise in the weights, which not exactly linear. This forces the weights to follow the prior and keeps the reconstruction robust to outliers.  
Importantly, the resulting components from the constrained NMF are physically meaningful with respect to the true function. In a crystallographic or spectroscopic context, this is important because peak intensity carries physical significance.

\begin{figure}
    \includegraphics{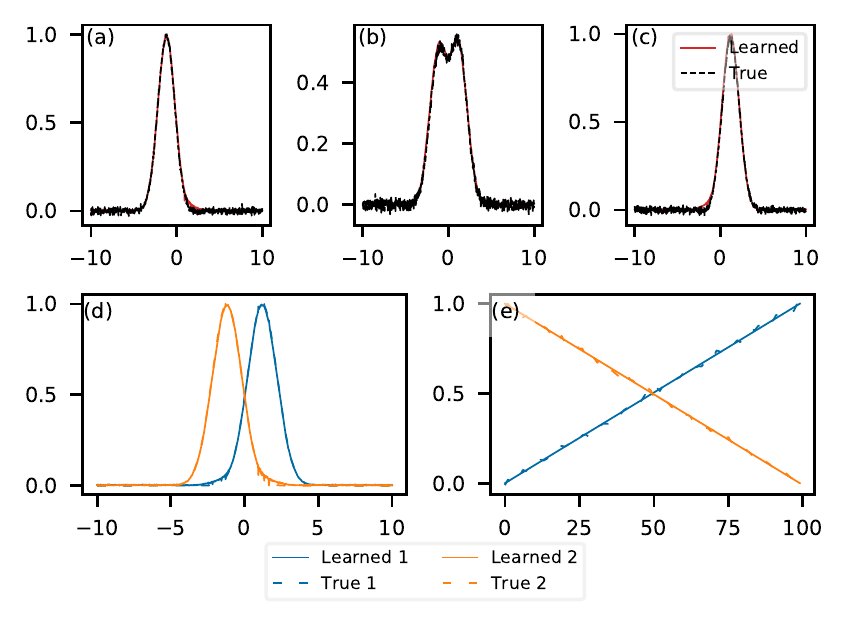}
    \caption{\label{fig:toy_weights} The reconstruction of the (a) first, (b) median, and (c) last pattern of a dataset of mixed Gaussian functions using constrained NMF. The weights of the decomposition were constrained to be linear with respect to the dataset index. The full dataset is shown in Figure~\ref{fig:datasets}(a). 
    (d) Both the ordering and amplitude of the learned components successfully matches the ground truth functions.
    (e) By fixing the weights, they cannot match the non-linearity in the ground truth but serve as a physically meaningful approximation.}
\end{figure}

The results from this constrained NMF are obtained rapidly, and can be produced on-the-fly during measurement. The output components offer immediate insight into the underlying dataset, while depending on an physical approximate prior. 
These components can be used as starting conditions for exact refinement, such as a Rietveld refinement,\cite{Rietveld_1967} of each pattern or spectrum in a dataset. 
When analysis is less time constrained, other methods for further tuning the optimization of constrained NMF can be explored with continual human input. Some such methods include using non-rigid constraints and Markov chain Monte Carlo optimization,\cite{Geddes_2019} which would enable a flexible, non-exact, linear prior. 

\subsection{\label{sec:toy_components}Constraining components with synthetic datasets}
As a second controlled example, we created a synthetic dataset from three diverse functions with smooth, non-monotonically varying weights in a linear combination (Fig~\ref{fig:datasets}b). Those functions included an overlapping box and Gaussian function, and a Lorentz function which does not overlap either.
In this scenario, knowledge of any of these components significantly constrains the NP-hard optimization problem of NMF. 
\par

\begin{figure}
    \includegraphics{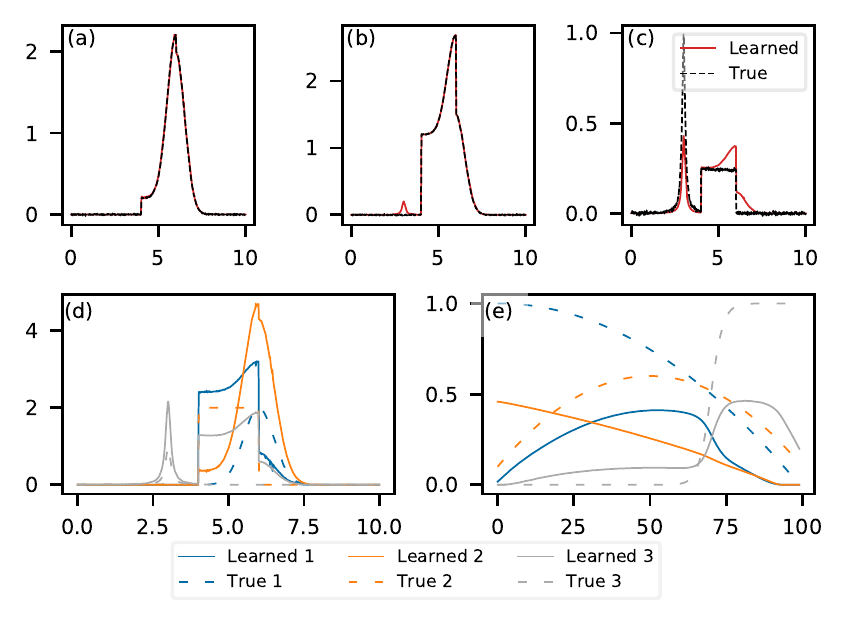}
    \caption{\label{fig:toy_funky}  The reconstruction of the (a) first, (b) median, and (c) last pattern of a dataset of mixed Gaussian, Lorenzian, and box functions using canonical NMF. The full dataset is shown in Figure~\ref{fig:datasets}(b). 
    (d) Canonical NMF fails to successfully decompose this dataset and produce an accurate reconstruction. The learned components are mixtures of the underlying functions.
    (e) The learned weights compensate for the inaccurate components, albeit the decomposition is trapped in a local optimum.}
\end{figure}

Unlike the previous example, this dataset was not effectively decomposed by canonical NMF. The resulting reconstruction MSE was 25.441 with a mean R-factor of 0.034.
One reason for this failure was the non-differentiable box function. This property makes decomposition particularly challenging for the gradient descent methods used in the optimization.
As shown in Figure~\ref{fig:toy_funky}d, the model found a local minimum in the optimization where the learned components were partial mixtures of the true components.
This is both physically inaccurate and results in a poor reconstruction.
In a case such as this, the decomposition from NMF results in a weak, non-physical, dimensionality reduction, which offers no new knowledge to the scientist. 
\par

\begin{figure}
    \includegraphics{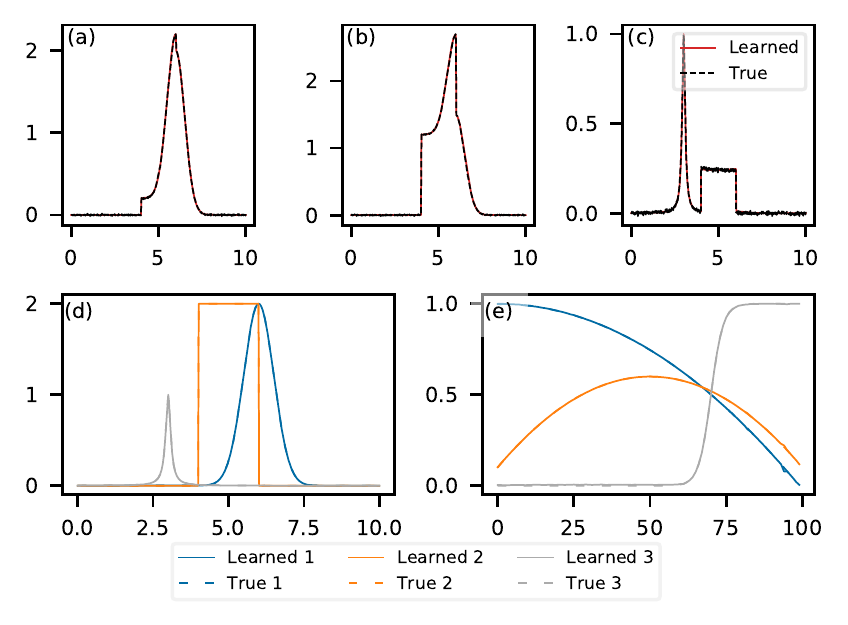}
    \caption{\label{fig:toy_components}The reconstruction of the (a) first, (b) median, and (c) last pattern of a dataset of mixed Gaussian, Lorenzian, and box functions using constrained NMF. The full dataset is shown in Figure~\ref{fig:datasets}(b). 
    (d) Prior knowledge of the functions fixes the underlying components.
    (e) The learned weights are meaningful and create an accurate reconstruction of the dataset. 
    }
\end{figure}

Corollary to the previous example where prior knowledge of the weights enabled a physically meaningful extraction of components, when the components are fed to NMF as a prior to decompose these composite functions, we were able to extract the ground truth weights.
As shown in Figure~\ref{fig:toy_funky}e, the weights are not trivial, and prior knowledge of the component functions enabled the model to learn those weights (Fig.~\ref{fig:toy_components}).
Again we see the rigidity of the priors ignoring the inherent noise, guiding the model away from overfitting. Yet, the constrained NMF outperformed canonical NMF even by the standard metrics (MSE=1.336, $<\mathrm{R-factor}>=$0.013). 
This approach is also applicable if only some of the components are known (Fig.~S1--6), albeit less robust for this use case.
This suggest that a procedure which allows for iterative inclusion of more priors through human or automatic intervention would allow for the most complex datasets to be successfully resolved using constrained NMF. 

\subsection{\label{sec:bto}Constraints enable physically meaningful phase segregation of BaTiO$_3$.}
Following this validation of constrained NMF in controlled systems, we sought to test its ability to construct a meaningful decomposition in a subtle and challenging crystallographic problem with measured data.
The BaTiO$_3$ dataset presented in Figure~\ref{fig:both_exp} contains four distinct phases across the temperature range measured. These transitions are subtle and not detectable considering traditional methods without expert intervention. Between these second order phase transitions, there is continuous peak shifting due to thermal expansion. 
\par

\begin{figure}
    \includegraphics{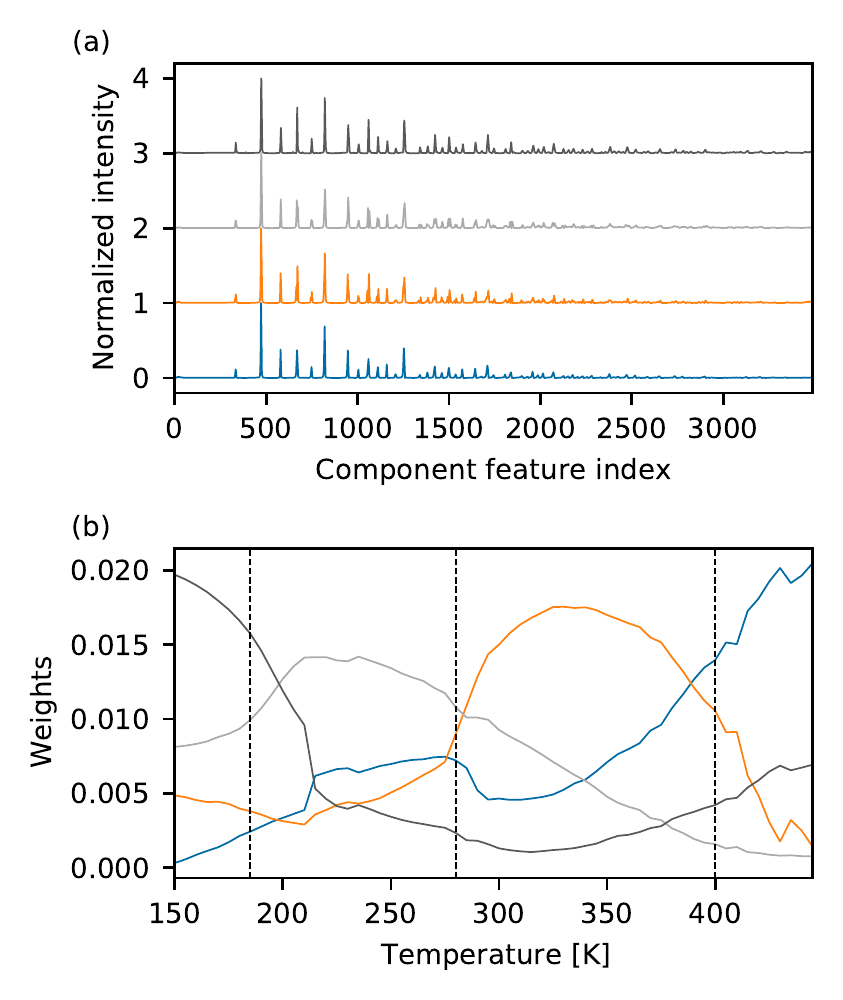}
    \caption{\label{fig:BTO} Canonical NMF fails to produce a physically meaningful decomposition of the variable temperature BaTiO$_3$ dataset (Fig.~\ref{fig:both_exp}(a)).
    (a) The learned components are randomized mixtures of real patterns.
    (b) The resulting weights carry no significance regarding the composition or any other property of the physical system. 
    }
\end{figure}

Canonical NMF is poorly suited to address translational variance (peak shifting) in decomposing a dataset into composite patterns. \cite{Stanev_2018, Bai_2018, Suram_2017} However, it was also incapable of identifying the clear second order phase transitions within the data.
Instead it learned mixtures of the true components as its pure components.
From Figure~\ref{fig:BTO}b, one might falsely conclude that the high temperature phase (blue) increases approximately linearly in composition with temperature, or that there are three components which make up the low temperature phase.
Despite this, the reconstruction loss from NMF was quite low with an MSE of 0.150 and a mean R-factor of 0.039. 
This demonstrates that canonical NMF is not well suited for crystallographic datasets with subtle transitions, a common case for diffraction studies.
\par

We next considered constraining the components of NMF to extract a better, more physically realistic, decomposition of the BaTiO$_3$ dataset. We sought to approach this in a way that made no presumption of a specific phase, and only operated under the knowledge that the dataset was measured along a single state variable (temperature).
Our synthetic experiment mixing diverse functions demonstrated that constraining all of the components led to the optimal solution, albeit partial constraints enabled the extraction of physically sensible intermediates. This concept has been demonstrated previously using different approaches on diffraction data.\cite{Gu_2021, Geddes_2019} 
To construct a physically meaningful solution without explicit knowledge of the underlying phases, we designed an iterative algorithm that wraps our constrained NMF to handle datasets over state variables such as temperature or composition.
\par

The iterative application of constrained NMF automatically selects components to constrain, thus depending only on the user selected number of components. 
It begins by assuming the end members of a dataset (in this case lowest and highest temperature patterns) are unique, fixing those members as components, and optimizing a partially constrained NMF. 
In the limiting case of a rank 2 NMF, this amounts to combinatorial decomposition, which has been successful in interpreting PDF data.\cite{Olds_2017}
When more components are desired, the algorithm chooses the most populated variable component, then finds the location where this component is most prevalent in the dataset. 
The next component is then fixed to the member at this location, and the NMF optimization proceeds with one additional constraint. This process continues until all members of the NMF are constrained. 
Critically for total scattering datasets, this results in weights that are bounded between 0 and 1, thus possessing a meaningful mapping to composition.
\par

\begin{figure}
    \includegraphics{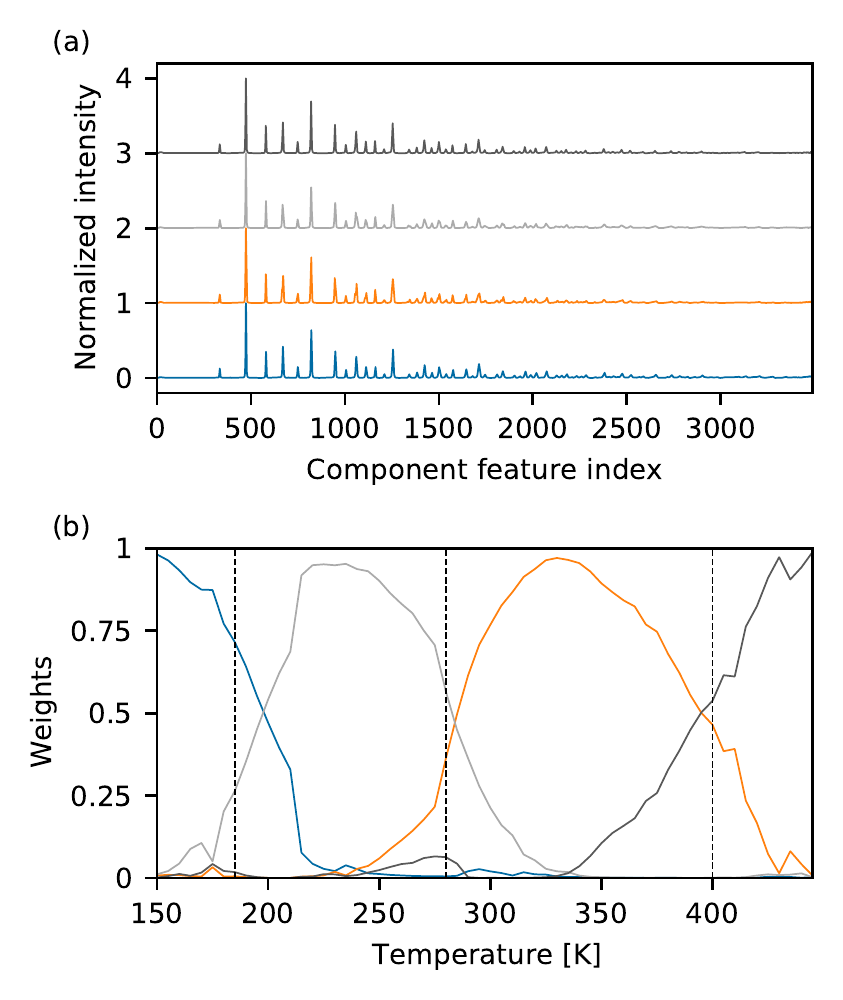}
    \caption{\label{fig:BTO_constrained} Automated constrained NMF produces a physically meaningful decomposition of the variable temperature BaTiO$_3$ dataset (Fig.~\ref{fig:both_exp}(a)).
    (a) The learned components correspond to the pure phases at the median temperatures where they are observed.
    (b) The resulting weights approximate mixing between phases and thermal expansion causing translational variance in the XRD patterns.}
\end{figure}

The automated, iterative, constrained NMF approached a similar reconstruction accuracy as the canonical NMF (MSE=0.349, $<\mathrm{R-factor}>=$0.051). However, the resulting components and weights are immediately interpretable by the scientist.
As the fixed components were chosen directly from the measured dataset, these can be used directly as reference diffraction patterns for pure phases (Fig.~\ref{fig:BTO_constrained}a).
Furthermore, the weights can be interpreted as molar fractions, showing phase mixtures during transitions (Fig.~\ref{fig:BTO_constrained}b). The applications of constraints --- even without prior knowledge of what those constraints should be --- enabled a substantial improvement over canonical NMF, providing both physically meaningful components and weights.
\par

There is still some room for improvement of this method. The transitions provided by the constrained NMF are still too gradual for second order phase transitions. While they occur in approximately the correct location, they indicate more mixing than is actually observed.
This is a direct result of translation dependence of NMF: peak shifts in pure phases are likely cast as mixtures between minimally and maximally shifted components that are otherwise identical. Supervised or semi-supervised convolutional deep learning methods have exceeded this performance.\cite{Oviedo_2019,Lee_2020, Maffettone_2021} Nonetheless, to the authors' knowledge this is the best performance of any fully unsupervised ML approach at segregating the phases in such a dataset.

\subsection{\label{sec:salt}Constraints enable physically meaningful \emph{in situ} analysis of melting salts.}
\begin{figure}
    \includegraphics{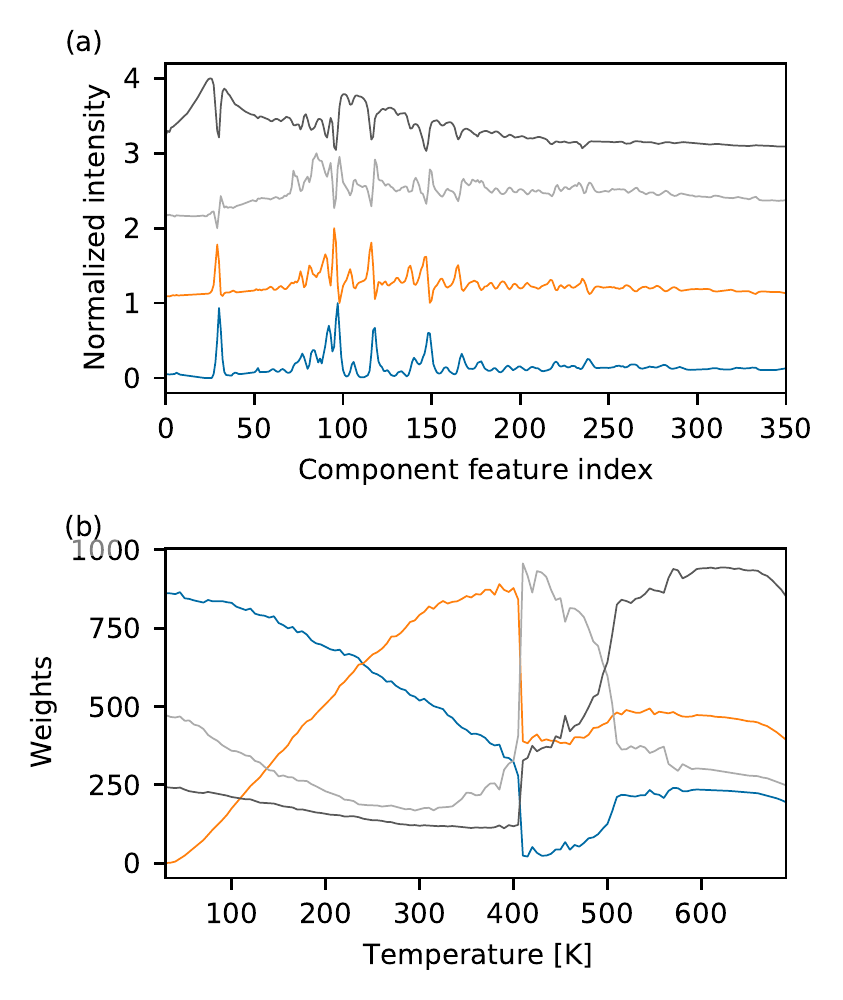}
    \caption{\label{fig:salt} Canonical NMF produces a confusing decomposition, offering limited clarity to the scientist.
    (a) The learned components are non-physical and cannot be interpreted as XRD patterns. 
    (b) The learned weights provide little information beyond an abrupt change around 400~\,K.
    }
\end{figure}

We then used the constrained NMF to monitor an experimental transition into an amorphous state with an unknown number of component phases. Canonical NMF behaves as expected with the same shortcomings we noted in the study of BaTiO$_3$. It elucidated the solid to amorphous phase boundary with a clear second order phase transition shown at 676\,K (Fig.~\ref{fig:salt}). The reconstruction MSE was 3.355  with a mean R-factor of 0.020. By integrating the NMF into data acquisition pipeline,\cite{Bluesky} this enables the researcher to note the transition temperature on-the-fly; however, the canonical approach is unable to extract meaningful components.
Without additional processing, the solid material that existed past the transition temperature was unable to be characterized.
During such an experiment---especially at a central facility---it is crucial to know whether the high temperature solid phase is unique and mandates further investigation at the time of measurement.
In the case of any experiment that is irreversible, destructive, or expensive to reproduce, immediate access to this information is paramount.
\par

\begin{figure}
    \includegraphics{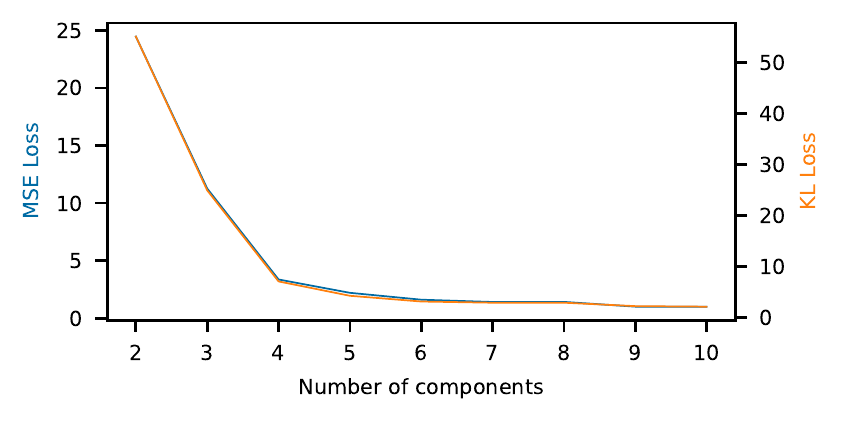}
    \caption{\label{fig:learning_curve} Following the loss curve of NMF with respect to the number of components shows diminishing returns around 4 components. Since this heuristic identification is subjective in nature, multiple instances of NMF can be run with a range of components.  }
\end{figure}

The constrained NMF approach offered this immediate analysis capability. Using the elbow method\cite{Thorndike_1953} to examine the MSE and KL-divergence loss, we noted diminishing returns in the NMF reconstruction at around 4 components (Fig.~\ref{fig:learning_curve}). The 4-component MSE was 9.193 with a mean R-factor of 0.0334.
Considering the region around the inflection point in these learning curves, the uncertain effects of peak shifting from thermal expansion, and potential secondary phases, we considered a constrained NMF that allowed for 3, 4, and 5 components.
Using the same automatic iterative constraining procedure from above, the constrained NMF elucidated a significantly more meaningful decomposition than canonical NMF.
\par

\begin{figure*}
    \includegraphics{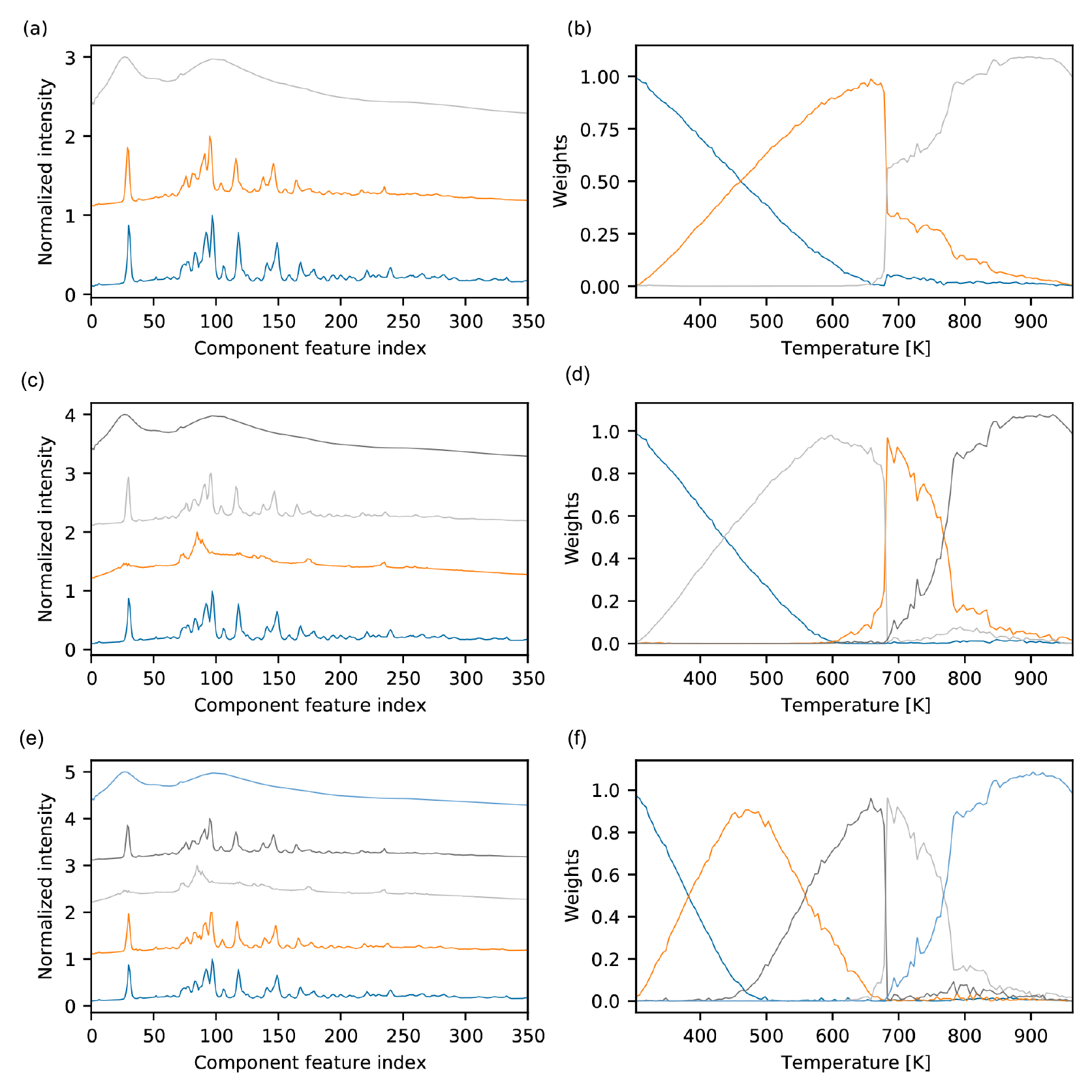}
    \caption{\label{fig:salt_summary} Automated constrained NMF was run using 3-6 components asynchronously on the molten salt dataset.
    The learned components for the (a) 3, (c) 4, and (e) 6 component decompostions provide insight into thermal expansion, a phases present during the temperature ramp. 
    The learned weights for the  (b) 3, (d) 4, and (f) 6 component decompositions demonstrate the second-order phase transition, the characteristic response to thermal expansion, and persistence of a solid phase after 400~\,K.
    }
\end{figure*}

Using multiple constrained NMFs of different rank elucidated the difference between a unique and prominent component, and a shifting component due to thermal expansion. In the rank-3 constrained NMF, two solid components and one amorphous component arose in the decomposition (Fig.~\ref{fig:salt_summary}a--b).
The solid components were nearly identical and showed a linear variation in weight (composition), which was 0.5/0.5 at the midpoint in the temperature range prior to the second order phase transition.
A small and decaying amount of the high temperature solid phase was present early after the transition. This linear exchange between the two similar components below 400\,K arises from the translational dependence of NMF: the algorithm compensates by considering two identical [besides shift] patterns at both ends of the relevant temperature range.
From the rank-3 constrained NMF alone, we can see the gradual impact of thermal expansion followed by a second order transition where some of the material remained solid.
\par

Considering higher orders of constrained NMF showed the uniqueness of the high temperature solid material, and further validated the assertion that smooth, linear variations of similar patterns are an indication of peak shifting.
The rank-4 NMF extracts similar components as the rank-3, albeit with a new high temperature component to describe the fine structure observed after the transition temperature (Fig.~\ref{fig:salt_summary}c--d).
This pattern (orange in Figure~\ref{fig:salt_summary}c) was a result of an amorphous background (later fully pronounced at high temperature in dark grey) and persistent, strained, crystallites of the original solid that had a preferred orientation with respect to the X-ray beam.
Lastly, the rank-6 NMF (Fig.~\ref{fig:salt_summary}e--f) decomposed the dataset into the same components with an additional subdivision of the low temperature solid phase. This subdivision occurred at the midpoint of the previous two end members, demonstrating the behavior of the constrained NMF when compensating for pattern shift.
This response to the imperceptible shift in the data is a useful and predictable benefit to the iterative and constrained approach: without translational invariance, this behavior enables the user to confidently note when components belong to the same, but shifting, phase. 

\par

The results of the constrained NMF for decomposing a variable temperature diffraction dataset are immediately more interpretable than a similar decomposition using cannonical NMF (Fig~\ref{fig:salt}). As with the BaTiO$_3$ example, the weights correspond to meaningful compositions.
Without solid-solid phase transitions as seen with BaTiO$_3$, the linear response to pattern shifting is apparent. While this ameliorates a problem with NMF, a more ideal solution would be one which maps only one component to a single phase regardless of the translational variations which occur during an experiment. 
Moreover, the recognition of persistent solids by the NMF is useful for adjusting an experimental protocol on the fly. Through user or algorithm intervention, the identification of a second order transition could be used to slow the rate of temperature change, allowing the system time to equilibrate.
In such experiments there is often a lag between the measured temperature and the material temperature, especially around a transition that involves latent heat. Indicating this transition on-the-fly enables a more robust and informative experiment.


\section{Conclusions} 
Here we developed a constrained NMF approach that interprets streaming spectral data on-the-fly, promptly providing critical insight and analysis from time-sensitive experiments such as XRD and PDF.
This approach efficiently handles the drastic increase in data generation rates --- critical at modern beamlines --- where conventional analysis approaches cannot keep up with the experimental pace required for optimized data collection.
We demonstrated the advantages of using constrained NMF to aid in revealing subtle phase transitions of a ferroelectric perovskite and the identification of experimental complexities in molten salts.  
This approach leverages the combined capabilities of the computer and the researcher.
Constrained NMF processes large datasets in an unsupervised fashion with an interactive prior: the intuition and decision making of the human-in-the-loop. 
Although applying constraints may cause the NMF model to create a worse reconstruction of the dataset, the additional prior information provided by constraints enables for a model that is more physically meaningful. 
These constraints can be chosen explicitly by the researcher, or by a implicitly from the dataset by an algorithm. 
This gives the streaming analysis added value that cannot be expected from canonical NMF.
Some overall limitations of NMF that still apply to constrained approaches are the inability to retain the partial translational invariance of XRD and PDF patterns,\cite{} as well as other common aberrations that impact a pattern without impacting the underlying phase.\cite{Oviedo_2019, Maffettone_2021} 
Constrained NMF can be extended to a broad range of spectral applications such as X-ray photoelectron spectra, X-ray absorption near edge spectra, photoluminescence spectra, nuclear magnetic resonance spectra, or mass spectra.\cite{Gu_2021}
The outputs can be interpreted asynchronously with supervised ML, and be used to inform adaptive experiments.\cite{Bluesky}
Constrained NMF enables human-in-the-loop, informed, unsupervised learning. It maintains the agency of the researcher through leveraging of expertise, while producing real-time analysis and actionable results during an experiment. 

\section{\label{sec:methods}Methods}
\subsection{NMF}
Given a non-negative data matrix $\textbf{X}$ of dimensions $m \times n$ --- where $m$ is the number of independent measurements and $n$ is the length of the feature dimension --- the goal of NMF is to find a factorization such that:
\begin{equation}
    \textbf{X} \approx \textbf{W}\textbf{H},
\end{equation}
where \textbf{W} and \textbf{H} are non-negative matrices of dimension $m \times k$ and $k \times n$, respectively, and $k$ is the user-specified number of components. As such, the problem reduces to solving:
\begin{equation}
    \min_{\textbf{W} \in \mathbb{R}^{m \times c}, \textbf{H} \in \mathbb{R}^{c \times n}} D(\textbf{X}, \textbf{W}\textbf{H}) \quad \text{s.t.} \quad \textbf{W} \geq 0, \textbf{H} \geq 0,
\end{equation}
where $D$ is a separable, positive measure of fit and the notation $\textbf{M} \geq 0$ indicates non-negativity of the entries of matrix $M$. 

We employ the $\beta$-divergence $D_\beta$ as a loss function during minimization. The $\beta$-divergence is a popular cost function for NMF minimization parameterized by a single shape parameter $\beta$ \citep{Fevotte_2011}. For the purposes of this study, we consider two special cases of the $\beta$-divergence --- squared Euclidean distance, also known as the squared error (SE) ($\beta=2$):
\begin{equation}
\label{eq:squared_error}
    D_2(\textbf{X}, \textbf{X}') = \frac{1}{2}\sum_{i=1}^{m}\sum_{j=1}^{n} (\textbf{X}_{i,j}-\textbf{X}'_{i,j})^2,
\end{equation}
and Kullback-Leibler divergence ($\beta=1$):
\begin{equation}
    D_1(\textbf{X}, \textbf{X}') = \sum_{i=1}^{m}\sum_{j=1}^{n} \left( \textbf{X}_{i,j} \log \frac{\textbf{X}_{i,j}}{\textbf{X}'_{i,j}} - \textbf{X}_{i,j} + \textbf{X}' _{i,j} \right),
\end{equation}
though the methods presented here may generalize to any choice of $\beta$.

We employ the alternating non-negative least squares (ANLS) algorithm in order to minimize the $\beta$-divergence subject to NMF constraints. Briefly, this algorithm operates by alternately minimizing $D_{\beta}(\textbf{X}, \textbf{WH})$ with respect to $\textbf{W}$ and $\textbf{H}$ under the constraints that all matrices remain non-negative. This can be solved by gradient descent with non-negativity constraints on parameter updates, as described in Algorithm \ref{alg:anls}:

\begin{algorithm}[H]
\caption{Alternating non-negative least squares}
\begin{algorithmic}[1]
\Procedure{ANLS}{$\textbf{X}$, $\textbf{W}_0$, $\textbf{H}_0$, $\beta$, $\lambda_1$, $\lambda_2$, iter, tol}
\State $i \gets 0$, \quad $L_{\text{prev}} \gets \infty$, \quad $\Delta \gets \infty$
\State $\textbf{W} \gets \textbf{W}_0$, \quad $\textbf{H} \gets \textbf{H}_0$
\While{$i \leq \text{iter}$ and $\Delta > \text{tol}$}
    \State $\textbf{X}' \gets$ \Call{Threshold}{$\textbf{WH}$, 0, $10^{-8}$}
    \State $L \gets D_{\beta}(\textbf{X}, \textbf{X}')$
    \State $\mu_{\textbf{W}} \gets \left\{ \begin{matrix}
      \sum_{j=1}^{n} \textbf{H}_{.,j} & \text{if } \beta=1 \\
      \textbf{X}'^{(\beta-1)} \textbf{W}^\intercal & \text{otherwise}
    \end{matrix}\right.\, $
    
    \State $\textbf{W} \gets \textbf{W} \cdot \max \left( \mu_{\textbf{W}} - \frac{\partial L}{\partial \textbf{W}}, 0 \right) / (\mu_{\textbf{W}} + \lambda_1 + \lambda_2 \textbf{W})$ \Comment{Non-negative $\textbf{W}$ update}
    
    \State $\textbf{X}' \gets$ \Call{Threshold}{$\textbf{WH}$, 0, $10^{-8}$}
    \State $L \gets D_{\beta}(\textbf{X}, \textbf{X}')$
    \State $\mu_{\textbf{H}} \gets \left\{ \begin{matrix}
      \sum_{i=1}^{m} \textbf{W}_{i,.} & \text{if } \beta=1 \\
      \textbf{W}^\intercal \textbf{X}'^{(\beta-1)} & \text{otherwise}
    \end{matrix}\right.\, $
    
    \State $\textbf{H} \gets \textbf{H} \cdot \max \left( \mu_{\textbf{H}} - \frac{\partial L}{\partial \textbf{H}}, 0 \right) / (\mu_{\textbf{H}} + \lambda_1 + \lambda_2 \textbf{H})$ \Comment{Non-negative $\textbf{H}$ update}
    \State $i \gets i+1$, \quad $\Delta \gets L-L_{\text{prev}}$, \quad $L_{\text{prev}} \gets L$
\EndWhile
\State \textbf{return} \textbf{W}, \textbf{H}
\EndProcedure

\Function{Threshold}{$M$, $t$, $\epsilon$}\Comment{Threshold each element in matrix $M$}
\For{all $M_{i,j}$ in $M$}
\State $ M_{i,j} \gets \left\{ \begin{matrix}
      M_{i,j} & \text{if } M_{i,j} > t \\
      \epsilon & \text{otherwise}
    \end{matrix}\right.\,$ 
\EndFor
\EndFunction

\end{algorithmic}
\label{alg:anls}
\end{algorithm}

To allow for import of prior knowledge, Algorithm \ref{alg:anls} allows for the user to supply initial values for $\textbf{W}$ and $\textbf{H}$ in the form of inputs $\textbf{W}_0$ and $\textbf{H}_0$. To fix certain columns of $\textbf{W}$ or rows of $\textbf{H}$ throughout optimization in order to reflect ground-truth knowledge, one may fix the gradients for known components to zero in the following manner:
\begin{equation}
    \frac{\partial D_{\beta}}{\partial \textbf{M}_{i,j}} = 0 \quad \forall \quad \textbf{M}_{i,j} \in S_\textbf{M},
\end{equation}
where $S_{\textbf{M}}$ is the set of all known components of matrix $\textbf{M}$.
\par
During the analysis of experimental results, we presented the MSE alongside the mean R-factor
\begin{equation}
\label{eq:R}
    R = \frac{\sum_{i=1}^{n}{|\textbf{X} - \textbf{WH}|}}
    {\sum_{i=1}^{n}{|\textbf{X}|}},
\end{equation}
as a commonly used crystallographic metric. 
\par
Algorithm \ref{alg:anls} has been implemented in PyTorch \cite{Paszke2019}, and is publicly available at \url{https://github.com/bnl/pub-Maffettone_2021_04}.
 
\subsection{Total scattering data collection}

Total scattering measurements were performed at the Pair-Distribution Funcion (PDF) beamline at the National Synchrotron Light Source II at Brookhaven National Laboratory.  The wavelength of incident X-rays for both measurements was found to be 0.1671 \AA \,(74.21 keV).  Data was collected in transmission mode with a flat-panel Perkin-Elmer detector.  Sample-to-detector distance and detector orientation was calibrated using Ni powder standard.  All raw 2D images were background subtracted using a dark image collected immediately prior to each scan.  Beamstop and detector aberrations were removed with masking prior to radial integration that was performed using pyFAI,\cite{ashiotis_2015} with PDFs then being generated using PDFgetX3.\cite{juhas_2013}

For the molten salt measurements, the sample-to-detector distance was found to be 282.74 mm.  The sample was held in 1 mm quartz capillaries resistevely heated using Nichrome wire, with temperatures measured via a K-type thermocouple.  Measurements were collected with 10 second data collection time per temperature.  For the BaTiO$_3$ measurments, samples were held in 1 mm quartz capillaries, with temperature controlled through a liquid nitrogen cryostream directed at the sample.  Diffraction (XRD) measurements were collected at 1 m from the sample with 60 second data collection time per temperature.  

\begin{acknowledgments}
We acknowledge financial support from the BNL Laboratory Directed Research and Development (LDRD) projects 20-032 “Accelerating materials discovery with total scattering via machine learning” (P.M.M., D.O.), and the Simons Foundation Center for Computational Biology (A.D). 
This research utilized the PDF (28-ID-1) Beamline and resources of the National Synchrotron Light Source II, a U.S. Department of Energy (DOE) Office of Science User Facility operated for the DOE Office of Science by Brookhaven National Laboratory under Contract No. DE-SC0012704.
\end{acknowledgments}

\section{Data Availability}
The data that support the findings of this study are available from the corresponding author upon reasonable request.

\bibliography{manuscript.bib}

\begin{thebibliography}{45}%
\makeatletter
\providecommand \@ifxundefined [1]{%
 \@ifx{#1\undefined}
}%
\providecommand \@ifnum [1]{%
 \ifnum #1\expandafter \@firstoftwo
 \else \expandafter \@secondoftwo
 \fi
}%
\providecommand \@ifx [1]{%
 \ifx #1\expandafter \@firstoftwo
 \else \expandafter \@secondoftwo
 \fi
}%
\providecommand \natexlab [1]{#1}%
\providecommand \enquote  [1]{``#1''}%
\providecommand \bibnamefont  [1]{#1}%
\providecommand \bibfnamefont [1]{#1}%
\providecommand \citenamefont [1]{#1}%
\providecommand \href@noop [0]{\@secondoftwo}%
\providecommand \href [0]{\begingroup \@sanitize@url \@href}%
\providecommand \@href[1]{\@@startlink{#1}\@@href}%
\providecommand \@@href[1]{\endgroup#1\@@endlink}%
\providecommand \@sanitize@url [0]{\catcode `\\12\catcode `\$12\catcode
  `\&12\catcode `\#12\catcode `\^12\catcode `\_12\catcode `\%12\relax}%
\providecommand \@@startlink[1]{}%
\providecommand \@@endlink[0]{}%
\providecommand \url  [0]{\begingroup\@sanitize@url \@url }%
\providecommand \@url [1]{\endgroup\@href {#1}{\urlprefix }}%
\providecommand \urlprefix  [0]{URL }%
\providecommand \Eprint [0]{\href }%
\providecommand \doibase [0]{http://dx.doi.org/}%
\providecommand \selectlanguage [0]{\@gobble}%
\providecommand \bibinfo  [0]{\@secondoftwo}%
\providecommand \bibfield  [0]{\@secondoftwo}%
\providecommand \translation [1]{[#1]}%
\providecommand \BibitemOpen [0]{}%
\providecommand \bibitemStop [0]{}%
\providecommand \bibitemNoStop [0]{.\EOS\space}%
\providecommand \EOS [0]{\spacefactor3000\relax}%
\providecommand \BibitemShut  [1]{\csname bibitem#1\endcsname}%
\let\auto@bib@innerbib\@empty
\bibitem [{\citenamefont {Kusne}\ \emph {et~al.}(2014)\citenamefont {Kusne},
  \citenamefont {Gao}, \citenamefont {Mehta}, \citenamefont {Ke}, \citenamefont
  {Nguyen}, \citenamefont {Ho}, \citenamefont {Antropov}, \citenamefont {Wang},
  \citenamefont {Kramer}, \citenamefont {Long},\ and\ \citenamefont
  {Takeuchi}}]{Kusne_2014}%
  \BibitemOpen
  \bibfield  {author} {\bibinfo {author} {\bibfnamefont {A.~G.}\ \bibnamefont
  {Kusne}}, \bibinfo {author} {\bibfnamefont {T.}~\bibnamefont {Gao}}, \bibinfo
  {author} {\bibfnamefont {A.}~\bibnamefont {Mehta}}, \bibinfo {author}
  {\bibfnamefont {L.}~\bibnamefont {Ke}}, \bibinfo {author} {\bibfnamefont
  {M.~C.}\ \bibnamefont {Nguyen}}, \bibinfo {author} {\bibfnamefont {K.-M.}\
  \bibnamefont {Ho}}, \bibinfo {author} {\bibfnamefont {V.}~\bibnamefont
  {Antropov}}, \bibinfo {author} {\bibfnamefont {C.-Z.}\ \bibnamefont {Wang}},
  \bibinfo {author} {\bibfnamefont {M.~J.}\ \bibnamefont {Kramer}}, \bibinfo
  {author} {\bibfnamefont {C.}~\bibnamefont {Long}}, \ and\ \bibinfo {author}
  {\bibfnamefont {I.}~\bibnamefont {Takeuchi}},\ }\bibfield  {title} {\enquote
  {\bibinfo {title} {On-the-fly machine-learning for high-throughput
  experiments: search for rare-earth-free permanent magnets},}\ }\href@noop {}
  {\bibfield  {journal} {\bibinfo  {journal} {Scientific Reports}\ }\textbf
  {\bibinfo {volume} {4}},\ \bibinfo {pages} {6367 EP --} (\bibinfo {year}
  {2014})}\BibitemShut {NoStop}%
\bibitem [{\citenamefont {G{\'o}mez-Bombarelli}\ \emph
  {et~al.}(2016)\citenamefont {G{\'o}mez-Bombarelli}, \citenamefont
  {Aguilera-Iparraguirre}, \citenamefont {Hirzel}, \citenamefont {Duvenaud},
  \citenamefont {Maclaurin}, \citenamefont {Blood-Forsythe}, \citenamefont
  {Chae}, \citenamefont {Einzinger}, \citenamefont {Ha}, \citenamefont {Wu},
  \citenamefont {Markopoulos}, \citenamefont {Jeon}, \citenamefont {Kang},
  \citenamefont {Miyazaki}, \citenamefont {Numata}, \citenamefont {Kim},
  \citenamefont {Huang}, \citenamefont {Hong}, \citenamefont {Baldo},
  \citenamefont {Adams},\ and\ \citenamefont
  {Aspuru-Guzik}}]{Gomez-Bombarelli_2016}%
  \BibitemOpen
  \bibfield  {author} {\bibinfo {author} {\bibfnamefont {R.}~\bibnamefont
  {G{\'o}mez-Bombarelli}}, \bibinfo {author} {\bibfnamefont {J.}~\bibnamefont
  {Aguilera-Iparraguirre}}, \bibinfo {author} {\bibfnamefont {T.~D.}\
  \bibnamefont {Hirzel}}, \bibinfo {author} {\bibfnamefont {D.}~\bibnamefont
  {Duvenaud}}, \bibinfo {author} {\bibfnamefont {D.}~\bibnamefont {Maclaurin}},
  \bibinfo {author} {\bibfnamefont {M.~A.}\ \bibnamefont {Blood-Forsythe}},
  \bibinfo {author} {\bibfnamefont {H.~S.}\ \bibnamefont {Chae}}, \bibinfo
  {author} {\bibfnamefont {M.}~\bibnamefont {Einzinger}}, \bibinfo {author}
  {\bibfnamefont {D.-G.}\ \bibnamefont {Ha}}, \bibinfo {author} {\bibfnamefont
  {T.}~\bibnamefont {Wu}}, \bibinfo {author} {\bibfnamefont {G.}~\bibnamefont
  {Markopoulos}}, \bibinfo {author} {\bibfnamefont {S.}~\bibnamefont {Jeon}},
  \bibinfo {author} {\bibfnamefont {H.}~\bibnamefont {Kang}}, \bibinfo {author}
  {\bibfnamefont {H.}~\bibnamefont {Miyazaki}}, \bibinfo {author}
  {\bibfnamefont {M.}~\bibnamefont {Numata}}, \bibinfo {author} {\bibfnamefont
  {S.}~\bibnamefont {Kim}}, \bibinfo {author} {\bibfnamefont {W.}~\bibnamefont
  {Huang}}, \bibinfo {author} {\bibfnamefont {S.~I.}\ \bibnamefont {Hong}},
  \bibinfo {author} {\bibfnamefont {M.}~\bibnamefont {Baldo}}, \bibinfo
  {author} {\bibfnamefont {R.~P.}\ \bibnamefont {Adams}}, \ and\ \bibinfo
  {author} {\bibfnamefont {A.}~\bibnamefont {Aspuru-Guzik}},\ }\bibfield
  {title} {\enquote {\bibinfo {title} {Design of efficient molecular organic
  light-emitting diodes by a high-throughput virtual screening and experimental
  approach},}\ }\href@noop {} {\bibfield  {journal} {\bibinfo  {journal} {Nat.
  Mater.}\ }\textbf {\bibinfo {volume} {15}},\ \bibinfo {pages} {1120 EP --}
  (\bibinfo {year} {2016})}\BibitemShut {NoStop}%
\bibitem [{\citenamefont {Greenaway}\ \emph {et~al.}(2018)\citenamefont
  {Greenaway}, \citenamefont {Santolini}, \citenamefont {Bennison},
  \citenamefont {Alston}, \citenamefont {Pugh}, \citenamefont {Little},
  \citenamefont {Miklitz}, \citenamefont {Eden-Rump}, \citenamefont {Clowes},
  \citenamefont {Shakil}, \citenamefont {Cuthbertson}, \citenamefont
  {Armstrong}, \citenamefont {Briggs}, \citenamefont {Jelfs},\ and\
  \citenamefont {Cooper}}]{Greenaway_2018}%
  \BibitemOpen
  \bibfield  {author} {\bibinfo {author} {\bibfnamefont {R.~L.}\ \bibnamefont
  {Greenaway}}, \bibinfo {author} {\bibfnamefont {V.}~\bibnamefont
  {Santolini}}, \bibinfo {author} {\bibfnamefont {M.~J.}\ \bibnamefont
  {Bennison}}, \bibinfo {author} {\bibfnamefont {B.~M.}\ \bibnamefont
  {Alston}}, \bibinfo {author} {\bibfnamefont {C.~J.}\ \bibnamefont {Pugh}},
  \bibinfo {author} {\bibfnamefont {M.~A.}\ \bibnamefont {Little}}, \bibinfo
  {author} {\bibfnamefont {M.}~\bibnamefont {Miklitz}}, \bibinfo {author}
  {\bibfnamefont {E.~G.~B.}\ \bibnamefont {Eden-Rump}}, \bibinfo {author}
  {\bibfnamefont {R.}~\bibnamefont {Clowes}}, \bibinfo {author} {\bibfnamefont
  {A.}~\bibnamefont {Shakil}}, \bibinfo {author} {\bibfnamefont {H.~J.}\
  \bibnamefont {Cuthbertson}}, \bibinfo {author} {\bibfnamefont
  {H.}~\bibnamefont {Armstrong}}, \bibinfo {author} {\bibfnamefont {M.~E.}\
  \bibnamefont {Briggs}}, \bibinfo {author} {\bibfnamefont {K.~E.}\
  \bibnamefont {Jelfs}}, \ and\ \bibinfo {author} {\bibfnamefont {A.~I.}\
  \bibnamefont {Cooper}},\ }\bibfield  {title} {\enquote {\bibinfo {title}
  {High-throughput discovery of organic cages and catenanes using computational
  screening fused with robotic synthesis},}\ }\href@noop {} {\bibfield
  {journal} {\bibinfo  {journal} {Nat. Commun.}\ }\textbf {\bibinfo {volume}
  {9}},\ \bibinfo {pages} {2849} (\bibinfo {year} {2018})}\BibitemShut
  {NoStop}%
\bibitem [{\citenamefont {Langner}\ \emph {et~al.}(2020)\citenamefont
  {Langner}, \citenamefont {H{\"a}se}, \citenamefont {Perea}, \citenamefont
  {Stubhan}, \citenamefont {Hauch}, \citenamefont {Roch}, \citenamefont
  {Heumueller}, \citenamefont {Aspuru-Guzik},\ and\ \citenamefont
  {Brabec}}]{Langner_2020}%
  \BibitemOpen
  \bibfield  {author} {\bibinfo {author} {\bibfnamefont {S.}~\bibnamefont
  {Langner}}, \bibinfo {author} {\bibfnamefont {F.}~\bibnamefont {H{\"a}se}},
  \bibinfo {author} {\bibfnamefont {J.~D.}\ \bibnamefont {Perea}}, \bibinfo
  {author} {\bibfnamefont {T.}~\bibnamefont {Stubhan}}, \bibinfo {author}
  {\bibfnamefont {J.}~\bibnamefont {Hauch}}, \bibinfo {author} {\bibfnamefont
  {L.~M.}\ \bibnamefont {Roch}}, \bibinfo {author} {\bibfnamefont
  {T.}~\bibnamefont {Heumueller}}, \bibinfo {author} {\bibfnamefont
  {A.}~\bibnamefont {Aspuru-Guzik}}, \ and\ \bibinfo {author} {\bibfnamefont
  {C.~J.}\ \bibnamefont {Brabec}},\ }\bibfield  {title} {\enquote {\bibinfo
  {title} {Beyond ternary opv: High-throughput experimentation and self-driving
  laboratories optimize multicomponent systems},}\ }\href@noop {} {\bibfield
  {journal} {\bibinfo  {journal} {Advanced Materials}\ }\textbf {\bibinfo
  {volume} {32}},\ \bibinfo {pages} {1907801} (\bibinfo {year}
  {2020})}\BibitemShut {NoStop}%
\bibitem [{\citenamefont {Ludwig}(2019)}]{Ludwig_2019}%
  \BibitemOpen
  \bibfield  {author} {\bibinfo {author} {\bibfnamefont {A.}~\bibnamefont
  {Ludwig}},\ }\bibfield  {title} {\enquote {\bibinfo {title} {Discovery of new
  materials using combinatorial synthesis and high-throughput characterization
  of thin-film materials libraries combined with computational methods},}\
  }\href@noop {} {\bibfield  {journal} {\bibinfo  {journal} {npj Comput.
  Mater.}\ }\textbf {\bibinfo {volume} {5}},\ \bibinfo {pages} {70} (\bibinfo
  {year} {2019})}\BibitemShut {NoStop}%
\bibitem [{\citenamefont {Decker}\ \emph {et~al.}(2017)\citenamefont {Decker},
  \citenamefont {Naujoks}, \citenamefont {Langenk{\"a}mper}, \citenamefont
  {Somsen},\ and\ \citenamefont {Ludwig}}]{Decker_2017}%
  \BibitemOpen
  \bibfield  {author} {\bibinfo {author} {\bibfnamefont {P.}~\bibnamefont
  {Decker}}, \bibinfo {author} {\bibfnamefont {D.}~\bibnamefont {Naujoks}},
  \bibinfo {author} {\bibfnamefont {D.}~\bibnamefont {Langenk{\"a}mper}},
  \bibinfo {author} {\bibfnamefont {C.}~\bibnamefont {Somsen}}, \ and\ \bibinfo
  {author} {\bibfnamefont {A.}~\bibnamefont {Ludwig}},\ }\bibfield  {title}
  {\enquote {\bibinfo {title} {High-throughput structural and functional
  characterization of the thin film materials system {Ni-Co-Al}},}\ }\href@noop
  {} {\bibfield  {journal} {\bibinfo  {journal} {ACS Comb. Sci.}\ }\textbf
  {\bibinfo {volume} {19}},\ \bibinfo {pages} {618--624} (\bibinfo {year}
  {2017})}\BibitemShut {NoStop}%
\bibitem [{\citenamefont {Yeung}\ \emph {et~al.}(2016)\citenamefont {Yeung},
  \citenamefont {Wu}, \citenamefont {Henke}, \citenamefont {Cheetham},
  \citenamefont {O'Hare},\ and\ \citenamefont {Walton}}]{Yeung_2016}%
  \BibitemOpen
  \bibfield  {author} {\bibinfo {author} {\bibfnamefont {H.~H.-M.}\
  \bibnamefont {Yeung}}, \bibinfo {author} {\bibfnamefont {Y.}~\bibnamefont
  {Wu}}, \bibinfo {author} {\bibfnamefont {S.}~\bibnamefont {Henke}}, \bibinfo
  {author} {\bibfnamefont {A.~K.}\ \bibnamefont {Cheetham}}, \bibinfo {author}
  {\bibfnamefont {D.}~\bibnamefont {O'Hare}}, \ and\ \bibinfo {author}
  {\bibfnamefont {R.~I.}\ \bibnamefont {Walton}},\ }\bibfield  {title}
  {\enquote {\bibinfo {title} {In-situ observation of successive
  crystallizations and metastable intermediates in the formation of
  metal--organic frameworks},}\ }\href@noop {} {\bibfield  {journal} {\bibinfo
  {journal} {Angewandte Chemie International Edition}\ }\textbf {\bibinfo
  {volume} {55}},\ \bibinfo {pages} {2012--2016} (\bibinfo {year}
  {2016})}\BibitemShut {NoStop}%
\bibitem [{\citenamefont {MacLeod}\ \emph {et~al.}(2020)\citenamefont
  {MacLeod}, \citenamefont {Parlane}, \citenamefont {Morrissey}, \citenamefont
  {H{\"a}se}, \citenamefont {Roch}, \citenamefont {Dettelbach}, \citenamefont
  {Moreira}, \citenamefont {Yunker}, \citenamefont {Rooney}, \citenamefont
  {Deeth}, \citenamefont {Lai}, \citenamefont {Ng}, \citenamefont {Situ},
  \citenamefont {Zhang}, \citenamefont {Elliott}, \citenamefont {Haley},
  \citenamefont {Dvorak}, \citenamefont {Aspuru-Guzik}, \citenamefont {Hein},\
  and\ \citenamefont {Berlinguette}}]{MacLeod_2020}%
  \BibitemOpen
  \bibfield  {author} {\bibinfo {author} {\bibfnamefont {B.~P.}\ \bibnamefont
  {MacLeod}}, \bibinfo {author} {\bibfnamefont {F.~G.~L.}\ \bibnamefont
  {Parlane}}, \bibinfo {author} {\bibfnamefont {T.~D.}\ \bibnamefont
  {Morrissey}}, \bibinfo {author} {\bibfnamefont {F.}~\bibnamefont {H{\"a}se}},
  \bibinfo {author} {\bibfnamefont {L.~M.}\ \bibnamefont {Roch}}, \bibinfo
  {author} {\bibfnamefont {K.~E.}\ \bibnamefont {Dettelbach}}, \bibinfo
  {author} {\bibfnamefont {R.}~\bibnamefont {Moreira}}, \bibinfo {author}
  {\bibfnamefont {L.~P.~E.}\ \bibnamefont {Yunker}}, \bibinfo {author}
  {\bibfnamefont {M.~B.}\ \bibnamefont {Rooney}}, \bibinfo {author}
  {\bibfnamefont {J.~R.}\ \bibnamefont {Deeth}}, \bibinfo {author}
  {\bibfnamefont {V.}~\bibnamefont {Lai}}, \bibinfo {author} {\bibfnamefont
  {G.~J.}\ \bibnamefont {Ng}}, \bibinfo {author} {\bibfnamefont
  {H.}~\bibnamefont {Situ}}, \bibinfo {author} {\bibfnamefont {R.~H.}\
  \bibnamefont {Zhang}}, \bibinfo {author} {\bibfnamefont {M.~S.}\ \bibnamefont
  {Elliott}}, \bibinfo {author} {\bibfnamefont {T.~H.}\ \bibnamefont {Haley}},
  \bibinfo {author} {\bibfnamefont {D.~J.}\ \bibnamefont {Dvorak}}, \bibinfo
  {author} {\bibfnamefont {A.}~\bibnamefont {Aspuru-Guzik}}, \bibinfo {author}
  {\bibfnamefont {J.~E.}\ \bibnamefont {Hein}}, \ and\ \bibinfo {author}
  {\bibfnamefont {C.~P.}\ \bibnamefont {Berlinguette}},\ }\bibfield  {title}
  {\enquote {\bibinfo {title} {Self-driving laboratory for accelerated
  discovery of thin-film materials},}\ }\href@noop {} {\bibfield  {journal}
  {\bibinfo  {journal} {Science Advances}\ }\textbf {\bibinfo {volume} {6}}
  (\bibinfo {year} {2020})}\BibitemShut {NoStop}%
\bibitem [{\citenamefont {Li}\ \emph {et~al.}(2020)\citenamefont {Li},
  \citenamefont {Najeeb}, \citenamefont {Alves}, \citenamefont {Sherman},
  \citenamefont {Shekar}, \citenamefont {Cruz~Parrilla}, \citenamefont
  {Pendleton}, \citenamefont {Wang}, \citenamefont {Nega}, \citenamefont
  {Zeller}, \citenamefont {Schrier}, \citenamefont {Norquist},\ and\
  \citenamefont {Chan}}]{Li_2020}%
  \BibitemOpen
  \bibfield  {author} {\bibinfo {author} {\bibfnamefont {Z.}~\bibnamefont
  {Li}}, \bibinfo {author} {\bibfnamefont {M.~A.}\ \bibnamefont {Najeeb}},
  \bibinfo {author} {\bibfnamefont {L.}~\bibnamefont {Alves}}, \bibinfo
  {author} {\bibfnamefont {A.~Z.}\ \bibnamefont {Sherman}}, \bibinfo {author}
  {\bibfnamefont {V.}~\bibnamefont {Shekar}}, \bibinfo {author} {\bibfnamefont
  {P.}~\bibnamefont {Cruz~Parrilla}}, \bibinfo {author} {\bibfnamefont {I.~M.}\
  \bibnamefont {Pendleton}}, \bibinfo {author} {\bibfnamefont {W.}~\bibnamefont
  {Wang}}, \bibinfo {author} {\bibfnamefont {P.~W.}\ \bibnamefont {Nega}},
  \bibinfo {author} {\bibfnamefont {M.}~\bibnamefont {Zeller}}, \bibinfo
  {author} {\bibfnamefont {J.}~\bibnamefont {Schrier}}, \bibinfo {author}
  {\bibfnamefont {A.~J.}\ \bibnamefont {Norquist}}, \ and\ \bibinfo {author}
  {\bibfnamefont {E.~M.}\ \bibnamefont {Chan}},\ }\bibfield  {title} {\enquote
  {\bibinfo {title} {Robot-accelerated perovskite investigation and
  discovery},}\ }\href@noop {} {\bibfield  {journal} {\bibinfo  {journal}
  {Chemistry of Materials}\ }\textbf {\bibinfo {volume} {32}},\ \bibinfo
  {pages} {5650--5663} (\bibinfo {year} {2020})}\BibitemShut {NoStop}%
\bibitem [{\citenamefont {Fleischer}\ \emph {et~al.}(2018)\citenamefont
  {Fleischer}, \citenamefont {Baumann}, \citenamefont {Joshi}, \citenamefont
  {Chu}, \citenamefont {Roddelkopf}, \citenamefont {Klos},\ and\ \citenamefont
  {Thurow}}]{Fleischer_2018}%
  \BibitemOpen
  \bibfield  {author} {\bibinfo {author} {\bibfnamefont {H.}~\bibnamefont
  {Fleischer}}, \bibinfo {author} {\bibfnamefont {D.}~\bibnamefont {Baumann}},
  \bibinfo {author} {\bibfnamefont {S.}~\bibnamefont {Joshi}}, \bibinfo
  {author} {\bibfnamefont {X.}~\bibnamefont {Chu}}, \bibinfo {author}
  {\bibfnamefont {T.}~\bibnamefont {Roddelkopf}}, \bibinfo {author}
  {\bibfnamefont {M.}~\bibnamefont {Klos}}, \ and\ \bibinfo {author}
  {\bibfnamefont {K.}~\bibnamefont {Thurow}},\ }\bibfield  {title} {\enquote
  {\bibinfo {title} {Analytical measurements and efficient process generation
  using a dual--arm robot equipped with electronic pipettes},}\ }\href@noop {}
  {\bibfield  {journal} {\bibinfo  {journal} {Energies}\ }\textbf {\bibinfo
  {volume} {11}} (\bibinfo {year} {2018})}\BibitemShut {NoStop}%
\bibitem [{\citenamefont {Campbell}\ \emph {et~al.}(2020)\citenamefont
  {Campbell}, \citenamefont {Allan}, \citenamefont {Barbour}, \citenamefont
  {Olds}, \citenamefont {Rakitin}, \citenamefont {Smith},\ and\ \citenamefont
  {Wilkins}}]{Campbell_2020}%
  \BibitemOpen
  \bibfield  {author} {\bibinfo {author} {\bibfnamefont {S.}~\bibnamefont
  {Campbell}}, \bibinfo {author} {\bibfnamefont {D.~B.}\ \bibnamefont {Allan}},
  \bibinfo {author} {\bibfnamefont {A.}~\bibnamefont {Barbour}}, \bibinfo
  {author} {\bibfnamefont {D.}~\bibnamefont {Olds}}, \bibinfo {author}
  {\bibfnamefont {M.}~\bibnamefont {Rakitin}}, \bibinfo {author} {\bibfnamefont
  {R.}~\bibnamefont {Smith}}, \ and\ \bibinfo {author} {\bibfnamefont {S.~B.}\
  \bibnamefont {Wilkins}},\ }\bibfield  {title} {\enquote {\bibinfo {title}
  {Outlook for artificial intelligence and machine learning at the nsls-ii},}\
  }\href@noop {} {\bibfield  {journal} {\bibinfo  {journal} {Machine Learning:
  Science and Technology}\ } (\bibinfo {year} {2020})}\BibitemShut {NoStop}%
\bibitem [{\citenamefont {Olds}\ \emph {et~al.}(2017)\citenamefont {Olds},
  \citenamefont {Peterson}, \citenamefont {Crawford}, \citenamefont {Neilson},
  \citenamefont {Wang}, \citenamefont {Whitfield},\ and\ \citenamefont
  {Page}}]{Olds_2017}%
  \BibitemOpen
  \bibfield  {author} {\bibinfo {author} {\bibfnamefont {D.}~\bibnamefont
  {Olds}}, \bibinfo {author} {\bibfnamefont {P.~F.}\ \bibnamefont {Peterson}},
  \bibinfo {author} {\bibfnamefont {M.~K.}\ \bibnamefont {Crawford}}, \bibinfo
  {author} {\bibfnamefont {J.~R.}\ \bibnamefont {Neilson}}, \bibinfo {author}
  {\bibfnamefont {H.-W.}\ \bibnamefont {Wang}}, \bibinfo {author}
  {\bibfnamefont {P.~S.}\ \bibnamefont {Whitfield}}, \ and\ \bibinfo {author}
  {\bibfnamefont {K.}~\bibnamefont {Page}},\ }\bibfield  {title} {\enquote
  {\bibinfo {title} {{Combinatorial appraisal of transition states for {\it in
  situ} pair distribution function analysis}},}\ }\href@noop {} {\bibfield
  {journal} {\bibinfo  {journal} {J. Appl. Cryst.}\ }\textbf {\bibinfo {volume}
  {50}},\ \bibinfo {pages} {1744--1753} (\bibinfo {year} {2017})}\BibitemShut
  {NoStop}%
\bibitem [{\citenamefont {Hua}\ \emph {et~al.}(2021)\citenamefont {Hua},
  \citenamefont {Eggeman}, \citenamefont {Castillo-Mart{\'\i}nez},
  \citenamefont {Robert}, \citenamefont {Geddes}, \citenamefont {Lu},
  \citenamefont {Pickard}, \citenamefont {Meng}, \citenamefont {Wiaderek},
  \citenamefont {Pereira}, \citenamefont {Amatucci}, \citenamefont {Midgley},
  \citenamefont {Chapman}, \citenamefont {Steiner}, \citenamefont {Goodwin},\
  and\ \citenamefont {Grey}}]{Hua_2021}%
  \BibitemOpen
  \bibfield  {author} {\bibinfo {author} {\bibfnamefont {X.}~\bibnamefont
  {Hua}}, \bibinfo {author} {\bibfnamefont {A.~S.}\ \bibnamefont {Eggeman}},
  \bibinfo {author} {\bibfnamefont {E.}~\bibnamefont {Castillo-Mart{\'\i}nez}},
  \bibinfo {author} {\bibfnamefont {R.}~\bibnamefont {Robert}}, \bibinfo
  {author} {\bibfnamefont {H.~S.}\ \bibnamefont {Geddes}}, \bibinfo {author}
  {\bibfnamefont {Z.}~\bibnamefont {Lu}}, \bibinfo {author} {\bibfnamefont
  {C.~J.}\ \bibnamefont {Pickard}}, \bibinfo {author} {\bibfnamefont
  {W.}~\bibnamefont {Meng}}, \bibinfo {author} {\bibfnamefont {K.~M.}\
  \bibnamefont {Wiaderek}}, \bibinfo {author} {\bibfnamefont {N.}~\bibnamefont
  {Pereira}}, \bibinfo {author} {\bibfnamefont {G.~G.}\ \bibnamefont
  {Amatucci}}, \bibinfo {author} {\bibfnamefont {P.~A.}\ \bibnamefont
  {Midgley}}, \bibinfo {author} {\bibfnamefont {K.~W.}\ \bibnamefont
  {Chapman}}, \bibinfo {author} {\bibfnamefont {U.}~\bibnamefont {Steiner}},
  \bibinfo {author} {\bibfnamefont {A.~L.}\ \bibnamefont {Goodwin}}, \ and\
  \bibinfo {author} {\bibfnamefont {C.~P.}\ \bibnamefont {Grey}},\ }\bibfield
  {title} {\enquote {\bibinfo {title} {Revisiting metal fluorides as
  lithium-ion battery cathodes},}\ }\href@noop {} {\bibfield  {journal}
  {\bibinfo  {journal} {Nature Materials}\ } (\bibinfo {year}
  {2021})}\BibitemShut {NoStop}%
\bibitem [{\citenamefont {Shadike}\ \emph {et~al.}(2021)\citenamefont
  {Shadike}, \citenamefont {Lee}, \citenamefont {Borodin}, \citenamefont {Cao},
  \citenamefont {Fan}, \citenamefont {Wang}, \citenamefont {Lin}, \citenamefont
  {Bak}, \citenamefont {Ghose}, \citenamefont {Xu}, \citenamefont {Wang},
  \citenamefont {Liu}, \citenamefont {Xiao}, \citenamefont {Yang},\ and\
  \citenamefont {Hu}}]{Shadike_2021}%
  \BibitemOpen
  \bibfield  {author} {\bibinfo {author} {\bibfnamefont {Z.}~\bibnamefont
  {Shadike}}, \bibinfo {author} {\bibfnamefont {H.}~\bibnamefont {Lee}},
  \bibinfo {author} {\bibfnamefont {O.}~\bibnamefont {Borodin}}, \bibinfo
  {author} {\bibfnamefont {X.}~\bibnamefont {Cao}}, \bibinfo {author}
  {\bibfnamefont {X.}~\bibnamefont {Fan}}, \bibinfo {author} {\bibfnamefont
  {X.}~\bibnamefont {Wang}}, \bibinfo {author} {\bibfnamefont {R.}~\bibnamefont
  {Lin}}, \bibinfo {author} {\bibfnamefont {S.-M.}\ \bibnamefont {Bak}},
  \bibinfo {author} {\bibfnamefont {S.}~\bibnamefont {Ghose}}, \bibinfo
  {author} {\bibfnamefont {K.}~\bibnamefont {Xu}}, \bibinfo {author}
  {\bibfnamefont {C.}~\bibnamefont {Wang}}, \bibinfo {author} {\bibfnamefont
  {J.}~\bibnamefont {Liu}}, \bibinfo {author} {\bibfnamefont {J.}~\bibnamefont
  {Xiao}}, \bibinfo {author} {\bibfnamefont {X.-Q.}\ \bibnamefont {Yang}}, \
  and\ \bibinfo {author} {\bibfnamefont {E.}~\bibnamefont {Hu}},\ }\bibfield
  {title} {\enquote {\bibinfo {title} {Identification of lih and
  nanocrystalline lif in the solid--electrolyte interphase of lithium metal
  anodes},}\ }\href@noop {} {\bibfield  {journal} {\bibinfo  {journal} {Nature
  Nanotechnology}\ } (\bibinfo {year} {2021})}\BibitemShut {NoStop}%
\bibitem [{\citenamefont {Wang}, \citenamefont {Li},\ and\ \citenamefont
  {Domen}(2019)}]{Wang_2019}%
  \BibitemOpen
  \bibfield  {author} {\bibinfo {author} {\bibfnamefont {Z.}~\bibnamefont
  {Wang}}, \bibinfo {author} {\bibfnamefont {C.}~\bibnamefont {Li}}, \ and\
  \bibinfo {author} {\bibfnamefont {K.}~\bibnamefont {Domen}},\ }\bibfield
  {title} {\enquote {\bibinfo {title} {Recent developments in heterogeneous
  photocatalysts for solar-driven overall water splitting},}\ }\href@noop {}
  {\bibfield  {journal} {\bibinfo  {journal} {Chem. Soc. Rev.}\ }\textbf
  {\bibinfo {volume} {48}},\ \bibinfo {pages} {2109--2125} (\bibinfo {year}
  {2019})}\BibitemShut {NoStop}%
\bibitem [{\citenamefont {L{\"o}ffler}\ \emph {et~al.}(2019)\citenamefont
  {L{\"o}ffler}, \citenamefont {Savan}, \citenamefont {Garz{\'o}n-Manj{\'o}n},
  \citenamefont {Meischein}, \citenamefont {Scheu}, \citenamefont {Ludwig},\
  and\ \citenamefont {Schuhmann}}]{Loffler_2019}%
  \BibitemOpen
  \bibfield  {author} {\bibinfo {author} {\bibfnamefont {T.}~\bibnamefont
  {L{\"o}ffler}}, \bibinfo {author} {\bibfnamefont {A.}~\bibnamefont {Savan}},
  \bibinfo {author} {\bibfnamefont {A.}~\bibnamefont {Garz{\'o}n-Manj{\'o}n}},
  \bibinfo {author} {\bibfnamefont {M.}~\bibnamefont {Meischein}}, \bibinfo
  {author} {\bibfnamefont {C.}~\bibnamefont {Scheu}}, \bibinfo {author}
  {\bibfnamefont {A.}~\bibnamefont {Ludwig}}, \ and\ \bibinfo {author}
  {\bibfnamefont {W.}~\bibnamefont {Schuhmann}},\ }\bibfield  {title} {\enquote
  {\bibinfo {title} {Toward a paradigm shift in electrocatalysis using complex
  solid solution nanoparticles},}\ }\href@noop {} {\bibfield  {journal}
  {\bibinfo  {journal} {ACS Energy Letters}\ }\textbf {\bibinfo {volume} {4}},\
  \bibinfo {pages} {1206--1214} (\bibinfo {year} {2019})}\BibitemShut {NoStop}%
\bibitem [{\citenamefont {Olds}(2020)}]{Olds_2020}%
  \BibitemOpen
  \bibfield  {author} {\bibinfo {author} {\bibfnamefont {D.}~\bibnamefont
  {Olds}},\ }\bibfield  {title} {\enquote {\bibinfo {title} {Synchrotron x-ray
  diffraction for energy and environmental materials: The current role and
  future directions of total scattering beamlines in the functional material
  scientific ecosystem},}\ }\href@noop {} {\bibfield  {journal} {\bibinfo
  {journal} {Synchrotron Radiation News}\ }\textbf {\bibinfo {volume} {33}},\
  \bibinfo {pages} {4--10} (\bibinfo {year} {2020})}\BibitemShut {NoStop}%
\bibitem [{\citenamefont {Clarke}\ \emph {et~al.}(2021)\citenamefont {Clarke},
  \citenamefont {Ablitt}, \citenamefont {Daniels}, \citenamefont {Checchia},\
  and\ \citenamefont {Senn}}]{Clarke_2021}%
  \BibitemOpen
  \bibfield  {author} {\bibinfo {author} {\bibfnamefont {G.}~\bibnamefont
  {Clarke}}, \bibinfo {author} {\bibfnamefont {C.}~\bibnamefont {Ablitt}},
  \bibinfo {author} {\bibfnamefont {J.}~\bibnamefont {Daniels}}, \bibinfo
  {author} {\bibfnamefont {S.}~\bibnamefont {Checchia}}, \ and\ \bibinfo
  {author} {\bibfnamefont {M.~S.}\ \bibnamefont {Senn}},\ }\bibfield  {title}
  {\enquote {\bibinfo {title} {{{\it In situ} X-ray diffraction investigation
  of electric-field-induced switching in a hybrid improper ferroelectric}},}\
  }\href@noop {} {\bibfield  {journal} {\bibinfo  {journal} {Journal of Applied
  Crystallography}\ }\textbf {\bibinfo {volume} {54}} (\bibinfo {year}
  {2021})}\BibitemShut {NoStop}%
\bibitem [{\citenamefont {Rietveld}(1967)}]{Rietveld_1967}%
  \BibitemOpen
  \bibfield  {author} {\bibinfo {author} {\bibfnamefont {H.~M.}\ \bibnamefont
  {Rietveld}},\ }\bibfield  {title} {\enquote {\bibinfo {title} {{Line profiles
  of neutron powder-diffraction peaks for structure refinement}},}\ }\href@noop
  {} {\bibfield  {journal} {\bibinfo  {journal} {Acta Cryst.}\ }\textbf
  {\bibinfo {volume} {22}},\ \bibinfo {pages} {151--152} (\bibinfo {year}
  {1967})}\BibitemShut {NoStop}%
\bibitem [{\citenamefont {Giacovazzo}(2011)}]{Giacovazzo_2011}%
  \BibitemOpen
  \bibinfo {editor} {\bibfnamefont {C.}~\bibnamefont {Giacovazzo}},\ ed.,\
  \href@noop {} {\emph {\bibinfo {title} {{Fundamentals of
  Crystallography}}}},\ \bibinfo {edition} {3rd}\ ed.\ (\bibinfo  {publisher}
  {Oxford University Press},\ \bibinfo {year} {2011})\BibitemShut {NoStop}%
\bibitem [{\citenamefont {Iwasaki}, \citenamefont {Kusne},\ and\ \citenamefont
  {Takeuchi}(2017)}]{Iwasaki_2017}%
  \BibitemOpen
  \bibfield  {author} {\bibinfo {author} {\bibfnamefont {Y.}~\bibnamefont
  {Iwasaki}}, \bibinfo {author} {\bibfnamefont {A.~G.}\ \bibnamefont {Kusne}},
  \ and\ \bibinfo {author} {\bibfnamefont {I.}~\bibnamefont {Takeuchi}},\
  }\bibfield  {title} {\enquote {\bibinfo {title} {Comparison of dissimilarity
  measures for cluster analysis of x-ray diffraction data from combinatorial
  libraries},}\ }\href@noop {} {\bibfield  {journal} {\bibinfo  {journal} {npj
  Comput. Mater.}\ }\textbf {\bibinfo {volume} {3}},\ \bibinfo {pages} {1--9}
  (\bibinfo {year} {2017})}\BibitemShut {NoStop}%
\bibitem [{\citenamefont {Xiong}\ \emph {et~al.}(2017)\citenamefont {Xiong},
  \citenamefont {He}, \citenamefont {Hattrick-Simpers},\ and\ \citenamefont
  {Hu}}]{Xiong_2017}%
  \BibitemOpen
  \bibfield  {author} {\bibinfo {author} {\bibfnamefont {Z.}~\bibnamefont
  {Xiong}}, \bibinfo {author} {\bibfnamefont {Y.}~\bibnamefont {He}}, \bibinfo
  {author} {\bibfnamefont {J.~R.}\ \bibnamefont {Hattrick-Simpers}}, \ and\
  \bibinfo {author} {\bibfnamefont {J.}~\bibnamefont {Hu}},\ }\bibfield
  {title} {\enquote {\bibinfo {title} {Automated phase segmentation for
  large-scale x-ray diffraction data using a graph-based phase segmentation
  (gphase) algorithm},}\ }\href@noop {} {\bibfield  {journal} {\bibinfo
  {journal} {ACS Comb. Sci.}\ }\textbf {\bibinfo {volume} {19}},\ \bibinfo
  {pages} {137--144} (\bibinfo {year} {2017})}\BibitemShut {NoStop}%
\bibitem [{\citenamefont {Long}\ \emph {et~al.}(2009)\citenamefont {Long},
  \citenamefont {Bunker}, \citenamefont {Li}, \citenamefont {Karen},\ and\
  \citenamefont {Takeuchi}}]{Long_2009}%
  \BibitemOpen
  \bibfield  {author} {\bibinfo {author} {\bibfnamefont {C.~J.}\ \bibnamefont
  {Long}}, \bibinfo {author} {\bibfnamefont {D.}~\bibnamefont {Bunker}},
  \bibinfo {author} {\bibfnamefont {X.}~\bibnamefont {Li}}, \bibinfo {author}
  {\bibfnamefont {V.~L.}\ \bibnamefont {Karen}}, \ and\ \bibinfo {author}
  {\bibfnamefont {I.}~\bibnamefont {Takeuchi}},\ }\bibfield  {title} {\enquote
  {\bibinfo {title} {Rapid identification of structural phases in combinatorial
  thin-film libraries using x-ray diffraction and non-negative matrix
  factorization},}\ }\href@noop {} {\bibfield  {journal} {\bibinfo  {journal}
  {Rev. Sci. Instrum.}\ }\textbf {\bibinfo {volume} {80}},\ \bibinfo {pages}
  {103902} (\bibinfo {year} {2009})}\BibitemShut {NoStop}%
\bibitem [{\citenamefont {Oviedo}\ \emph {et~al.}(2019)\citenamefont {Oviedo},
  \citenamefont {Ren}, \citenamefont {Sun}, \citenamefont {Settens},
  \citenamefont {Liu}, \citenamefont {Hartono}, \citenamefont {Ramasamy},
  \citenamefont {DeCost}, \citenamefont {Tian},\ and\ \citenamefont
  {Romano}}]{Oviedo_2019}%
  \BibitemOpen
  \bibfield  {author} {\bibinfo {author} {\bibfnamefont {F.}~\bibnamefont
  {Oviedo}}, \bibinfo {author} {\bibfnamefont {Z.}~\bibnamefont {Ren}},
  \bibinfo {author} {\bibfnamefont {S.}~\bibnamefont {Sun}}, \bibinfo {author}
  {\bibfnamefont {C.}~\bibnamefont {Settens}}, \bibinfo {author} {\bibfnamefont
  {Z.}~\bibnamefont {Liu}}, \bibinfo {author} {\bibfnamefont {N.~T.~P.}\
  \bibnamefont {Hartono}}, \bibinfo {author} {\bibfnamefont {S.}~\bibnamefont
  {Ramasamy}}, \bibinfo {author} {\bibfnamefont {B.~L.}\ \bibnamefont
  {DeCost}}, \bibinfo {author} {\bibfnamefont {S.~I.~P.}\ \bibnamefont {Tian}},
  \ and\ \bibinfo {author} {\bibfnamefont {G.}~\bibnamefont {Romano}},\
  }\bibfield  {title} {\enquote {\bibinfo {title} {Fast and interpretable
  classification of small x-ray diffraction datasets using data augmentation
  and deep neural networks},}\ }\href@noop {} {\bibfield  {journal} {\bibinfo
  {journal} {npj Comput. Mater.}\ }\textbf {\bibinfo {volume} {5}},\ \bibinfo
  {pages} {1--9} (\bibinfo {year} {2019})}\BibitemShut {NoStop}%
\bibitem [{\citenamefont {Lee}\ \emph {et~al.}(2020)\citenamefont {Lee},
  \citenamefont {Park}, \citenamefont {Lee}, \citenamefont {Singh},\ and\
  \citenamefont {Sohn}}]{Lee_2020}%
  \BibitemOpen
  \bibfield  {author} {\bibinfo {author} {\bibfnamefont {J.-W.}\ \bibnamefont
  {Lee}}, \bibinfo {author} {\bibfnamefont {W.~B.}\ \bibnamefont {Park}},
  \bibinfo {author} {\bibfnamefont {J.~H.}\ \bibnamefont {Lee}}, \bibinfo
  {author} {\bibfnamefont {S.~P.}\ \bibnamefont {Singh}}, \ and\ \bibinfo
  {author} {\bibfnamefont {K.-S.}\ \bibnamefont {Sohn}},\ }\bibfield  {title}
  {\enquote {\bibinfo {title} {A deep-learning technique for phase
  identification in multiphase inorganic compounds using synthetic xrd powder
  patterns},}\ }\href@noop {} {\bibfield  {journal} {\bibinfo  {journal} {Nat.
  Commun.}\ }\textbf {\bibinfo {volume} {11}},\ \bibinfo {pages} {86} (\bibinfo
  {year} {2020})}\BibitemShut {NoStop}%
\bibitem [{\citenamefont {Ziletti}\ \emph {et~al.}(2018)\citenamefont
  {Ziletti}, \citenamefont {Kumar}, \citenamefont {Scheffler},\ and\
  \citenamefont {Ghiringhelli}}]{Ziletti_2018}%
  \BibitemOpen
  \bibfield  {author} {\bibinfo {author} {\bibfnamefont {A.}~\bibnamefont
  {Ziletti}}, \bibinfo {author} {\bibfnamefont {D.}~\bibnamefont {Kumar}},
  \bibinfo {author} {\bibfnamefont {M.}~\bibnamefont {Scheffler}}, \ and\
  \bibinfo {author} {\bibfnamefont {L.~M.}\ \bibnamefont {Ghiringhelli}},\
  }\bibfield  {title} {\enquote {\bibinfo {title} {Insightful classification of
  crystal structures using deep learning},}\ }\href@noop {} {\bibfield
  {journal} {\bibinfo  {journal} {Nat. Commun.}\ }\textbf {\bibinfo {volume}
  {9}},\ \bibinfo {pages} {2775} (\bibinfo {year} {2018})}\BibitemShut
  {NoStop}%
\bibitem [{\citenamefont {Aguiar}\ \emph {et~al.}(2019)\citenamefont {Aguiar},
  \citenamefont {Gong}, \citenamefont {Unocic}, \citenamefont {Tasdizen},\ and\
  \citenamefont {Miller}}]{Aguiar_2019}%
  \BibitemOpen
  \bibfield  {author} {\bibinfo {author} {\bibfnamefont {J.~A.}\ \bibnamefont
  {Aguiar}}, \bibinfo {author} {\bibfnamefont {M.~L.}\ \bibnamefont {Gong}},
  \bibinfo {author} {\bibfnamefont {R.~R.}\ \bibnamefont {Unocic}}, \bibinfo
  {author} {\bibfnamefont {T.}~\bibnamefont {Tasdizen}}, \ and\ \bibinfo
  {author} {\bibfnamefont {B.~D.}\ \bibnamefont {Miller}},\ }\bibfield  {title}
  {\enquote {\bibinfo {title} {Decoding crystallography from high-resolution
  electron imaging and diffraction datasets with deep learning},}\ }\href@noop
  {} {\bibfield  {journal} {\bibinfo  {journal} {Sci. Adv.}\ }\textbf {\bibinfo
  {volume} {5}} (\bibinfo {year} {2019})}\BibitemShut {NoStop}%
\bibitem [{\citenamefont {{Di Chen}}\ \emph {et~al.}(2020)\citenamefont {{Di
  Chen}}, \citenamefont {Bai}, \citenamefont {Zhao}, \citenamefont {Ament},
  \citenamefont {Gregoire},\ and\ \citenamefont {Gomes}}]{Chen_2020}%
  \BibitemOpen
  \bibfield  {author} {\bibinfo {author} {\bibnamefont {{Di Chen}}}, \bibinfo
  {author} {\bibfnamefont {Y.}~\bibnamefont {Bai}}, \bibinfo {author}
  {\bibfnamefont {W.}~\bibnamefont {Zhao}}, \bibinfo {author} {\bibfnamefont
  {S.}~\bibnamefont {Ament}}, \bibinfo {author} {\bibfnamefont {J.~M.}\
  \bibnamefont {Gregoire}}, \ and\ \bibinfo {author} {\bibfnamefont {C.~P.}\
  \bibnamefont {Gomes}},\ }\bibfield  {title} {\enquote {\bibinfo {title} {Deep
  reasoning networks for unsupervised pattern de-mixing with constraint
  reasoning},}\ }in\ \href@noop {} {\emph {\bibinfo {booktitle} {Proceedings of
  the 37th International Conference on Machine Learning}}},\ \bibinfo {series}
  {Proceedings of Machine Learning Research}, Vol.\ \bibinfo {volume} {119},\
  \bibinfo {editor} {edited by\ \bibinfo {editor} {\bibfnamefont
  {F.}~\bibnamefont {Bach}}\ and\ \bibinfo {editor} {\bibfnamefont
  {D.}~\bibnamefont {Blei}}}\ (\bibinfo  {publisher} {PMLR},\ \bibinfo {year}
  {2020})\BibitemShut {NoStop}%
\bibitem [{\citenamefont {Takeuchi}\ \emph {et~al.}(2005)\citenamefont
  {Takeuchi}, \citenamefont {Long}, \citenamefont {Famodu}, \citenamefont
  {Murakami}, \citenamefont {Hattrick-Simpers}, \citenamefont {Rubloff},
  \citenamefont {Stukowski},\ and\ \citenamefont {Rajan}}]{Takeuchi_2005}%
  \BibitemOpen
  \bibfield  {author} {\bibinfo {author} {\bibfnamefont {I.}~\bibnamefont
  {Takeuchi}}, \bibinfo {author} {\bibfnamefont {C.~J.}\ \bibnamefont {Long}},
  \bibinfo {author} {\bibfnamefont {O.~O.}\ \bibnamefont {Famodu}}, \bibinfo
  {author} {\bibfnamefont {M.}~\bibnamefont {Murakami}}, \bibinfo {author}
  {\bibfnamefont {J.}~\bibnamefont {Hattrick-Simpers}}, \bibinfo {author}
  {\bibfnamefont {G.~W.}\ \bibnamefont {Rubloff}}, \bibinfo {author}
  {\bibfnamefont {M.}~\bibnamefont {Stukowski}}, \ and\ \bibinfo {author}
  {\bibfnamefont {K.}~\bibnamefont {Rajan}},\ }\bibfield  {title} {\enquote
  {\bibinfo {title} {Data management and visualization of x-ray diffraction
  spectra from thin film ternary composition spreads},}\ }\href@noop {}
  {\bibfield  {journal} {\bibinfo  {journal} {Rev. Sci. Instrum.}\ }\textbf
  {\bibinfo {volume} {76}},\ \bibinfo {pages} {062223} (\bibinfo {year}
  {2005})}\BibitemShut {NoStop}%
\bibitem [{\citenamefont {Stanev}\ \emph {et~al.}(2018)\citenamefont {Stanev},
  \citenamefont {Vesselinov}, \citenamefont {Kusne}, \citenamefont
  {Antoszewski}, \citenamefont {Takeuchi},\ and\ \citenamefont
  {Alexandrov}}]{Stanev_2018}%
  \BibitemOpen
  \bibfield  {author} {\bibinfo {author} {\bibfnamefont {V.}~\bibnamefont
  {Stanev}}, \bibinfo {author} {\bibfnamefont {V.~V.}\ \bibnamefont
  {Vesselinov}}, \bibinfo {author} {\bibfnamefont {A.~G.}\ \bibnamefont
  {Kusne}}, \bibinfo {author} {\bibfnamefont {G.}~\bibnamefont {Antoszewski}},
  \bibinfo {author} {\bibfnamefont {I.}~\bibnamefont {Takeuchi}}, \ and\
  \bibinfo {author} {\bibfnamefont {B.~S.}\ \bibnamefont {Alexandrov}},\
  }\bibfield  {title} {\enquote {\bibinfo {title} {Unsupervised phase mapping
  of x-ray diffraction data by nonnegative matrix factorization integrated with
  custom clustering},}\ }\href@noop {} {\bibfield  {journal} {\bibinfo
  {journal} {npj Computational Materials}\ }\textbf {\bibinfo {volume} {4}},\
  \bibinfo {pages} {43} (\bibinfo {year} {2018})}\BibitemShut {NoStop}%
\bibitem [{\citenamefont {Bai}\ \emph {et~al.}(2017)\citenamefont {Bai},
  \citenamefont {Bjorck}, \citenamefont {Xue}, \citenamefont {Suram},
  \citenamefont {Gregoire},\ and\ \citenamefont {Gomes}}]{Bai_2017}%
  \BibitemOpen
  \bibfield  {author} {\bibinfo {author} {\bibfnamefont {J.}~\bibnamefont
  {Bai}}, \bibinfo {author} {\bibfnamefont {J.}~\bibnamefont {Bjorck}},
  \bibinfo {author} {\bibfnamefont {Y.}~\bibnamefont {Xue}}, \bibinfo {author}
  {\bibfnamefont {S.~K.}\ \bibnamefont {Suram}}, \bibinfo {author}
  {\bibfnamefont {J.}~\bibnamefont {Gregoire}}, \ and\ \bibinfo {author}
  {\bibfnamefont {C.}~\bibnamefont {Gomes}},\ }\bibfield  {title} {\enquote
  {\bibinfo {title} {Relaxation methods for constrained matrix factorization
  problems: Solving the phase mapping problem in materials discovery},}\ }in\
  \href@noop {} {\emph {\bibinfo {booktitle} {Integration of AI and OR
  Techniques in Constraint Programming}}},\ \bibinfo {editor} {edited by\
  \bibinfo {editor} {\bibfnamefont {D.}~\bibnamefont {Salvagnin}}\ and\
  \bibinfo {editor} {\bibfnamefont {M.}~\bibnamefont {Lombardi}}}\ (\bibinfo
  {publisher} {Springer International Publishing},\ \bibinfo {address} {Cham},\
  \bibinfo {year} {2017})\ pp.\ \bibinfo {pages} {104--112}\BibitemShut
  {NoStop}%
\bibitem [{\citenamefont {Bai}\ \emph {et~al.}(2018)\citenamefont {Bai},
  \citenamefont {Xue}, \citenamefont {Bjorck}, \citenamefont {Le~Bras},
  \citenamefont {Rappazzo}, \citenamefont {Bernstein}, \citenamefont {Suram},
  \citenamefont {van Dover}, \citenamefont {Gregoire},\ and\ \citenamefont
  {Gomes}}]{Bai_2018}%
  \BibitemOpen
  \bibfield  {author} {\bibinfo {author} {\bibfnamefont {J.}~\bibnamefont
  {Bai}}, \bibinfo {author} {\bibfnamefont {Y.}~\bibnamefont {Xue}}, \bibinfo
  {author} {\bibfnamefont {J.}~\bibnamefont {Bjorck}}, \bibinfo {author}
  {\bibfnamefont {R.}~\bibnamefont {Le~Bras}}, \bibinfo {author} {\bibfnamefont
  {B.}~\bibnamefont {Rappazzo}}, \bibinfo {author} {\bibfnamefont
  {R.}~\bibnamefont {Bernstein}}, \bibinfo {author} {\bibfnamefont {S.~K.}\
  \bibnamefont {Suram}}, \bibinfo {author} {\bibfnamefont {R.~B.}\ \bibnamefont
  {van Dover}}, \bibinfo {author} {\bibfnamefont {J.~M.}\ \bibnamefont
  {Gregoire}}, \ and\ \bibinfo {author} {\bibfnamefont {C.~P.}\ \bibnamefont
  {Gomes}},\ }\bibfield  {title} {\enquote {\bibinfo {title} {Phase mapper:
  Accelerating materials discovery with ai},}\ }\href@noop {} {\bibfield
  {journal} {\bibinfo  {journal} {AI Magazine}\ }\textbf {\bibinfo {volume}
  {39}},\ \bibinfo {pages} {15--26} (\bibinfo {year} {2018})}\BibitemShut
  {NoStop}%
\bibitem [{\citenamefont {Suram}\ \emph {et~al.}(2017)\citenamefont {Suram},
  \citenamefont {Xue}, \citenamefont {Bai}, \citenamefont {Le~Bras},
  \citenamefont {Rappazzo}, \citenamefont {Bernstein}, \citenamefont {Bjorck},
  \citenamefont {Zhou}, \citenamefont {van Dover}, \citenamefont {Gomes},\ and\
  \citenamefont {Gregoire}}]{Suram_2017}%
  \BibitemOpen
  \bibfield  {author} {\bibinfo {author} {\bibfnamefont {S.~K.}\ \bibnamefont
  {Suram}}, \bibinfo {author} {\bibfnamefont {Y.}~\bibnamefont {Xue}}, \bibinfo
  {author} {\bibfnamefont {J.}~\bibnamefont {Bai}}, \bibinfo {author}
  {\bibfnamefont {R.}~\bibnamefont {Le~Bras}}, \bibinfo {author} {\bibfnamefont
  {B.}~\bibnamefont {Rappazzo}}, \bibinfo {author} {\bibfnamefont
  {R.}~\bibnamefont {Bernstein}}, \bibinfo {author} {\bibfnamefont
  {J.}~\bibnamefont {Bjorck}}, \bibinfo {author} {\bibfnamefont
  {L.}~\bibnamefont {Zhou}}, \bibinfo {author} {\bibfnamefont {R.~B.}\
  \bibnamefont {van Dover}}, \bibinfo {author} {\bibfnamefont {C.~P.}\
  \bibnamefont {Gomes}}, \ and\ \bibinfo {author} {\bibfnamefont {J.~M.}\
  \bibnamefont {Gregoire}},\ }\bibfield  {title} {\enquote {\bibinfo {title}
  {Automated phase mapping with agilefd and its application to light absorber
  discovery in the v--mn--nb oxide system},}\ }\href@noop {} {\bibfield
  {journal} {\bibinfo  {journal} {ACS Combinatorial Science}\ }\textbf
  {\bibinfo {volume} {19}},\ \bibinfo {pages} {37--46} (\bibinfo {year}
  {2017})}\BibitemShut {NoStop}%
\bibitem [{\citenamefont {Geddes}\ \emph {et~al.}(2019)\citenamefont {Geddes},
  \citenamefont {Blade}, \citenamefont {McCabe}, \citenamefont {Hughes},\ and\
  \citenamefont {Goodwin}}]{Geddes_2019}%
  \BibitemOpen
  \bibfield  {author} {\bibinfo {author} {\bibfnamefont {H.~S.}\ \bibnamefont
  {Geddes}}, \bibinfo {author} {\bibfnamefont {H.}~\bibnamefont {Blade}},
  \bibinfo {author} {\bibfnamefont {J.~F.}\ \bibnamefont {McCabe}}, \bibinfo
  {author} {\bibfnamefont {L.~P.}\ \bibnamefont {Hughes}}, \ and\ \bibinfo
  {author} {\bibfnamefont {A.~L.}\ \bibnamefont {Goodwin}},\ }\bibfield
  {title} {\enquote {\bibinfo {title} {Structural characterisation of amorphous
  solid dispersions via metropolis matrix factorisation of pair distribution
  function data},}\ }\href@noop {} {\bibfield  {journal} {\bibinfo  {journal}
  {Chem. Commun.}\ }\textbf {\bibinfo {volume} {55}},\ \bibinfo {pages}
  {13346--13349} (\bibinfo {year} {2019})}\BibitemShut {NoStop}%
\bibitem [{\citenamefont {Li}\ \emph {et~al.}(2021)\citenamefont {Li},
  \citenamefont {Sprouster}, \citenamefont {Zheng}, \citenamefont {Neuefeind},
  \citenamefont {Braatz}, \citenamefont {Mcfarlane}, \citenamefont {Olds},
  \citenamefont {Lam}, \citenamefont {Li},\ and\ \citenamefont
  {Khaykovich}}]{Li_2021}%
  \BibitemOpen
  \bibfield  {author} {\bibinfo {author} {\bibfnamefont {Q.-J.}\ \bibnamefont
  {Li}}, \bibinfo {author} {\bibfnamefont {D.}~\bibnamefont {Sprouster}},
  \bibinfo {author} {\bibfnamefont {G.}~\bibnamefont {Zheng}}, \bibinfo
  {author} {\bibfnamefont {J.~C.}\ \bibnamefont {Neuefeind}}, \bibinfo {author}
  {\bibfnamefont {A.~D.}\ \bibnamefont {Braatz}}, \bibinfo {author}
  {\bibfnamefont {J.}~\bibnamefont {Mcfarlane}}, \bibinfo {author}
  {\bibfnamefont {D.}~\bibnamefont {Olds}}, \bibinfo {author} {\bibfnamefont
  {S.}~\bibnamefont {Lam}}, \bibinfo {author} {\bibfnamefont {J.}~\bibnamefont
  {Li}}, \ and\ \bibinfo {author} {\bibfnamefont {B.}~\bibnamefont
  {Khaykovich}},\ }\bibfield  {title} {\enquote {\bibinfo {title} {Complex
  structure of molten nacl--crcl3 salt: Cr--cl octahedral network and
  intermediate-range order},}\ }\href@noop {} {\bibfield  {journal} {\bibinfo
  {journal} {ACS Applied Energy Materials}\ } (\bibinfo {year}
  {2021})}\BibitemShut {NoStop}%
\bibitem [{\citenamefont {Paszke}\ \emph {et~al.}(2019)\citenamefont {Paszke},
  \citenamefont {Gross}, \citenamefont {Massa}, \citenamefont {Lerer},
  \citenamefont {Bradbury}, \citenamefont {Chanan}, \citenamefont {Killeen},
  \citenamefont {Lin}, \citenamefont {Gimelshein}, \citenamefont {Antiga},
  \citenamefont {Desmaison}, \citenamefont {Kopf}, \citenamefont {Yang},
  \citenamefont {DeVito}, \citenamefont {Raison}, \citenamefont {Tejani},
  \citenamefont {Chilamkurthy}, \citenamefont {Steiner}, \citenamefont {Fang},
  \citenamefont {Bai},\ and\ \citenamefont {Chintala}}]{Paszke2019}%
  \BibitemOpen
  \bibfield  {author} {\bibinfo {author} {\bibfnamefont {A.}~\bibnamefont
  {Paszke}}, \bibinfo {author} {\bibfnamefont {S.}~\bibnamefont {Gross}},
  \bibinfo {author} {\bibfnamefont {F.}~\bibnamefont {Massa}}, \bibinfo
  {author} {\bibfnamefont {A.}~\bibnamefont {Lerer}}, \bibinfo {author}
  {\bibfnamefont {J.}~\bibnamefont {Bradbury}}, \bibinfo {author}
  {\bibfnamefont {G.}~\bibnamefont {Chanan}}, \bibinfo {author} {\bibfnamefont
  {T.}~\bibnamefont {Killeen}}, \bibinfo {author} {\bibfnamefont
  {Z.}~\bibnamefont {Lin}}, \bibinfo {author} {\bibfnamefont {N.}~\bibnamefont
  {Gimelshein}}, \bibinfo {author} {\bibfnamefont {L.}~\bibnamefont {Antiga}},
  \bibinfo {author} {\bibfnamefont {A.}~\bibnamefont {Desmaison}}, \bibinfo
  {author} {\bibfnamefont {A.}~\bibnamefont {Kopf}}, \bibinfo {author}
  {\bibfnamefont {E.}~\bibnamefont {Yang}}, \bibinfo {author} {\bibfnamefont
  {Z.}~\bibnamefont {DeVito}}, \bibinfo {author} {\bibfnamefont
  {M.}~\bibnamefont {Raison}}, \bibinfo {author} {\bibfnamefont
  {A.}~\bibnamefont {Tejani}}, \bibinfo {author} {\bibfnamefont
  {S.}~\bibnamefont {Chilamkurthy}}, \bibinfo {author} {\bibfnamefont
  {B.}~\bibnamefont {Steiner}}, \bibinfo {author} {\bibfnamefont
  {L.}~\bibnamefont {Fang}}, \bibinfo {author} {\bibfnamefont {J.}~\bibnamefont
  {Bai}}, \ and\ \bibinfo {author} {\bibfnamefont {S.}~\bibnamefont
  {Chintala}},\ }\bibfield  {title} {\enquote {\bibinfo {title} {Pytorch: An
  imperative style, high-performance deep learning library},}\ }in\ \href@noop
  {} {\emph {\bibinfo {booktitle} {Advances in Neural Information Processing
  Systems 32}}},\ \bibinfo {editor} {edited by\ \bibinfo {editor}
  {\bibfnamefont {H.}~\bibnamefont {Wallach}}, \bibinfo {editor} {\bibfnamefont
  {H.}~\bibnamefont {Larochelle}}, \bibinfo {editor} {\bibfnamefont
  {A.}~\bibnamefont {Beygelzimer}}, \bibinfo {editor} {\bibfnamefont
  {F.}~\bibnamefont {d'Alch\'{e} Buc}}, \bibinfo {editor} {\bibfnamefont
  {E.}~\bibnamefont {Fox}}, \ and\ \bibinfo {editor} {\bibfnamefont
  {R.}~\bibnamefont {Garnett}}}\ (\bibinfo  {publisher} {Curran Associates,
  Inc.},\ \bibinfo {year} {2019})\ pp.\ \bibinfo {pages}
  {8024--8035}\BibitemShut {NoStop}%
\bibitem [{\citenamefont {Page}\ \emph {et~al.}(2010)\citenamefont {Page},
  \citenamefont {Proffen}, \citenamefont {Niederberger},\ and\ \citenamefont
  {Seshadri}}]{Page_2010}%
  \BibitemOpen
  \bibfield  {author} {\bibinfo {author} {\bibfnamefont {K.}~\bibnamefont
  {Page}}, \bibinfo {author} {\bibfnamefont {T.}~\bibnamefont {Proffen}},
  \bibinfo {author} {\bibfnamefont {M.}~\bibnamefont {Niederberger}}, \ and\
  \bibinfo {author} {\bibfnamefont {R.}~\bibnamefont {Seshadri}},\ }\bibfield
  {title} {\enquote {\bibinfo {title} {Probing local dipoles and ligand
  structure in batio3 nanoparticles},}\ }\href@noop {} {\bibfield  {journal}
  {\bibinfo  {journal} {Chemistry of Materials}\ }\textbf {\bibinfo {volume}
  {22}},\ \bibinfo {pages} {4386--4391} (\bibinfo {year} {2010})}\BibitemShut
  {NoStop}%
\bibitem [{\citenamefont {Wegner}\ \emph {et~al.}(2020)\citenamefont {Wegner},
  \citenamefont {Gu}, \citenamefont {James},\ and\ \citenamefont
  {Quandt}}]{Wegner_2020}%
  \BibitemOpen
  \bibfield  {author} {\bibinfo {author} {\bibfnamefont {M.}~\bibnamefont
  {Wegner}}, \bibinfo {author} {\bibfnamefont {H.}~\bibnamefont {Gu}}, \bibinfo
  {author} {\bibfnamefont {R.~D.}\ \bibnamefont {James}}, \ and\ \bibinfo
  {author} {\bibfnamefont {E.}~\bibnamefont {Quandt}},\ }\bibfield  {title}
  {\enquote {\bibinfo {title} {Correlation between phase compatibility and
  efficient energy conversion in zr-doped barium titanate},}\ }\href@noop {}
  {\bibfield  {journal} {\bibinfo  {journal} {Scientific Reports}\ }\textbf
  {\bibinfo {volume} {10}},\ \bibinfo {pages} {3496} (\bibinfo {year}
  {2020})}\BibitemShut {NoStop}%
\bibitem [{\citenamefont {Gu}, \citenamefont {Du},\ and\ \citenamefont
  {Billinge}(2021)}]{Gu_2021}%
  \BibitemOpen
  \bibfield  {author} {\bibinfo {author} {\bibfnamefont {R.}~\bibnamefont
  {Gu}}, \bibinfo {author} {\bibfnamefont {Q.}~\bibnamefont {Du}}, \ and\
  \bibinfo {author} {\bibfnamefont {S.~J.~L.}\ \bibnamefont {Billinge}},\
  }\href@noop {} {\enquote {\bibinfo {title} {A fast two-stage algorithm for
  non-negative matrix factorization in streaming data},}\ } (\bibinfo {year}
  {2021})\BibitemShut {NoStop}%
\bibitem [{\citenamefont {Maffettone}\ \emph {et~al.}(2021)\citenamefont
  {Maffettone}, \citenamefont {Banko}, \citenamefont {Cui}, \citenamefont
  {Lysogorskiy}, \citenamefont {Little}, \citenamefont {Olds}, \citenamefont
  {Ludwig},\ and\ \citenamefont {Cooper}}]{Maffettone_2021}%
  \BibitemOpen
  \bibfield  {author} {\bibinfo {author} {\bibfnamefont {P.~M.}\ \bibnamefont
  {Maffettone}}, \bibinfo {author} {\bibfnamefont {L.}~\bibnamefont {Banko}},
  \bibinfo {author} {\bibfnamefont {P.}~\bibnamefont {Cui}}, \bibinfo {author}
  {\bibfnamefont {Y.}~\bibnamefont {Lysogorskiy}}, \bibinfo {author}
  {\bibfnamefont {M.~A.}\ \bibnamefont {Little}}, \bibinfo {author}
  {\bibfnamefont {D.}~\bibnamefont {Olds}}, \bibinfo {author} {\bibfnamefont
  {A.}~\bibnamefont {Ludwig}}, \ and\ \bibinfo {author} {\bibfnamefont {A.~I.}\
  \bibnamefont {Cooper}},\ }\href@noop {} {\enquote {\bibinfo {title}
  {Crystallography companion agent for high-throughput materials discovery},}\
  } (\bibinfo {year} {2021})\BibitemShut {NoStop}%
\bibitem [{\citenamefont {Allan}\ \emph {et~al.}(2019)\citenamefont {Allan},
  \citenamefont {Caswell}, \citenamefont {Campbell},\ and\ \citenamefont
  {Rakitin}}]{Bluesky}%
  \BibitemOpen
  \bibfield  {author} {\bibinfo {author} {\bibfnamefont {D.}~\bibnamefont
  {Allan}}, \bibinfo {author} {\bibfnamefont {T.}~\bibnamefont {Caswell}},
  \bibinfo {author} {\bibfnamefont {S.}~\bibnamefont {Campbell}}, \ and\
  \bibinfo {author} {\bibfnamefont {M.}~\bibnamefont {Rakitin}},\ }\bibfield
  {title} {\enquote {\bibinfo {title} {Bluesky's ahead: A multi-facility
  collaboration for an a la carte software project for data acquisition and
  management},}\ }\href@noop {} {\bibfield  {journal} {\bibinfo  {journal}
  {Synchrotron Radiation News}\ }\textbf {\bibinfo {volume} {32}},\ \bibinfo
  {pages} {19--22} (\bibinfo {year} {2019})}\BibitemShut {NoStop}%
\bibitem [{\citenamefont {Thorndike}(1953)}]{Thorndike_1953}%
  \BibitemOpen
  \bibfield  {author} {\bibinfo {author} {\bibfnamefont {R.~L.}\ \bibnamefont
  {Thorndike}},\ }\bibfield  {title} {\enquote {\bibinfo {title} {Who belongs
  in the family?}}\ }\href@noop {} {\bibfield  {journal} {\bibinfo  {journal}
  {Psychometrika}\ }\textbf {\bibinfo {volume} {18}},\ \bibinfo {pages}
  {267--276} (\bibinfo {year} {1953})}\BibitemShut {NoStop}%
\bibitem [{\citenamefont {Févotte}\ and\ \citenamefont
  {Idier}(2011)}]{Fevotte_2011}%
  \BibitemOpen
  \bibfield  {author} {\bibinfo {author} {\bibfnamefont {C.}~\bibnamefont
  {Févotte}}\ and\ \bibinfo {author} {\bibfnamefont {J.}~\bibnamefont
  {Idier}},\ }\bibfield  {title} {\enquote {\bibinfo {title} {Algorithms for
  nonnegative matrix factorization with the beta-divergence},}\ }\href
  {\doibase 10.1162/NECO_a_00168} {\bibfield  {journal} {\bibinfo  {journal}
  {Neural Computation}\ }\textbf {\bibinfo {volume} {23}},\ \bibinfo {pages}
  {2421--2456} (\bibinfo {year} {2011})}\BibitemShut {NoStop}%
\bibitem [{\citenamefont {Ashiotis}\ \emph {et~al.}(2015)\citenamefont
  {Ashiotis}, \citenamefont {Deschildre}, \citenamefont {Nawaz}, \citenamefont
  {Wright}, \citenamefont {Karkoulis}, \citenamefont {Picca},\ and\
  \citenamefont {Kieffer}}]{ashiotis_2015}%
  \BibitemOpen
  \bibfield  {author} {\bibinfo {author} {\bibfnamefont {G.}~\bibnamefont
  {Ashiotis}}, \bibinfo {author} {\bibfnamefont {A.}~\bibnamefont
  {Deschildre}}, \bibinfo {author} {\bibfnamefont {Z.}~\bibnamefont {Nawaz}},
  \bibinfo {author} {\bibfnamefont {J.~P.}\ \bibnamefont {Wright}}, \bibinfo
  {author} {\bibfnamefont {D.}~\bibnamefont {Karkoulis}}, \bibinfo {author}
  {\bibfnamefont {F.~E.}\ \bibnamefont {Picca}}, \ and\ \bibinfo {author}
  {\bibfnamefont {J.}~\bibnamefont {Kieffer}},\ }\bibfield  {title} {\enquote
  {\bibinfo {title} {The fast azimuthal integration python library: pyfai},}\
  }\href@noop {} {\bibfield  {journal} {\bibinfo  {journal} {Journal of applied
  crystallography}\ }\textbf {\bibinfo {volume} {48}},\ \bibinfo {pages}
  {510--519} (\bibinfo {year} {2015})}\BibitemShut {NoStop}%
\bibitem [{\citenamefont {Juh{\'a}s}\ \emph {et~al.}(2013)\citenamefont
  {Juh{\'a}s}, \citenamefont {Davis}, \citenamefont {Farrow},\ and\
  \citenamefont {Billinge}}]{juhas_2013}%
  \BibitemOpen
  \bibfield  {author} {\bibinfo {author} {\bibfnamefont {P.}~\bibnamefont
  {Juh{\'a}s}}, \bibinfo {author} {\bibfnamefont {T.}~\bibnamefont {Davis}},
  \bibinfo {author} {\bibfnamefont {C.~L.}\ \bibnamefont {Farrow}}, \ and\
  \bibinfo {author} {\bibfnamefont {S.~J.}\ \bibnamefont {Billinge}},\
  }\bibfield  {title} {\enquote {\bibinfo {title} {Pdfgetx3: a rapid and highly
  automatable program for processing powder diffraction data into total
  scattering pair distribution functions},}\ }\href@noop {} {\bibfield
  {journal} {\bibinfo  {journal} {Journal of Applied Crystallography}\ }\textbf
  {\bibinfo {volume} {46}},\ \bibinfo {pages} {560--566} (\bibinfo {year}
  {2013})}\BibitemShut {NoStop}%
\end{thebibliography}%

\end{document}



\title[]{Supplementary Information:\\Constrained non-negative matrix factorization enabling real-time insights of \emph{in situ} and high-throughput experiments}

\author{Phillip M. Maffettone}
\affiliation{National Synchrotron Light Source II, Brookhaven National Laboratory, Upton, NY 11973, United States of America}
\author{Aidan C. Daly}%
\affiliation{ 
Center for Computational Biology, Flatiron Institute, U.S.A
}%

\author{Daniel Olds}
 \email{dolds@bnl.gov}
\affiliation{%
Brookhaven National Laboratory, Upton, NY 11973, United States of America}%

\date{\today}

\maketitle

\section{Partially constraining components with synthetic datasets}
When considering the dataset shown in the main text Figure~1b, knowing even some of the underlying components increases the effectiveness of the NMF decomposition. 

\begin{figure}
    \includegraphics{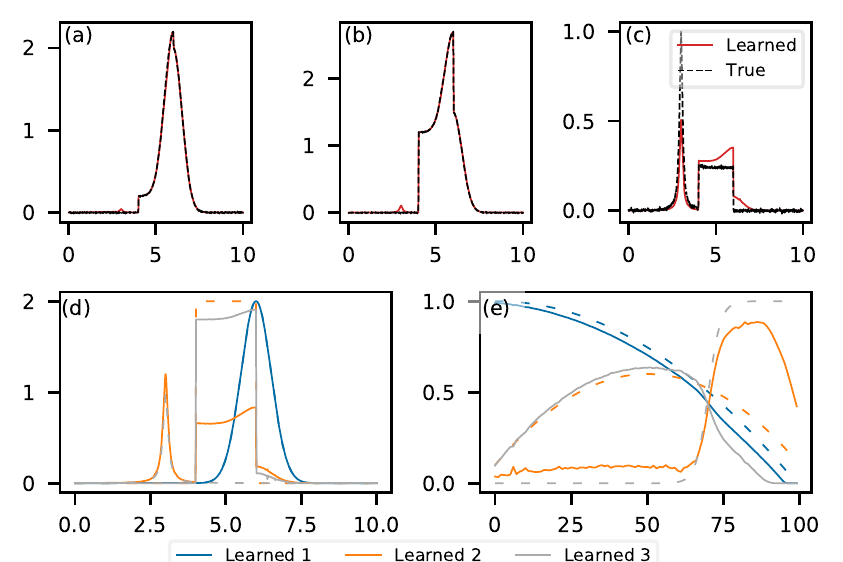}
    \caption{\label{fig:partial_0} The reconstruction of the (a) first, (b) median, and (c) last pattern of a dataset of mixed Gaussian, Lorenzian, and box functions using constrained NMF.
    (d) Prior knowledge of the Gaussian function fixes one underlying component.
    (e) The learned weights are improved and create an better reconstruction of the dataset, than canonical NMF (MSE=13.314, R-factor=0.029).
    }
\end{figure}

\begin{figure}
    \includegraphics{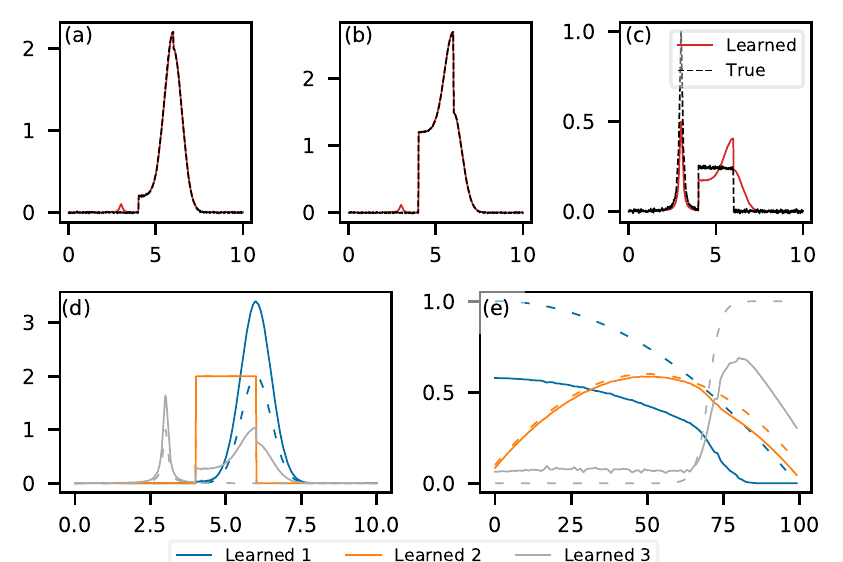}
    \caption{\label{fig:partial_1} The reconstruction of the (a) first, (b) median, and (c) last pattern of a dataset of mixed Gaussian, Lorenzian, and box functions using constrained NMF.
    (d) Prior knowledge of the box function fixes one underlying component.
    (e) The learned weights are improved and create an better reconstruction of the dataset, than canonical NMF (MSE=27.959, R-factor=0.044).
    }
\end{figure}

\begin{figure}
    \includegraphics{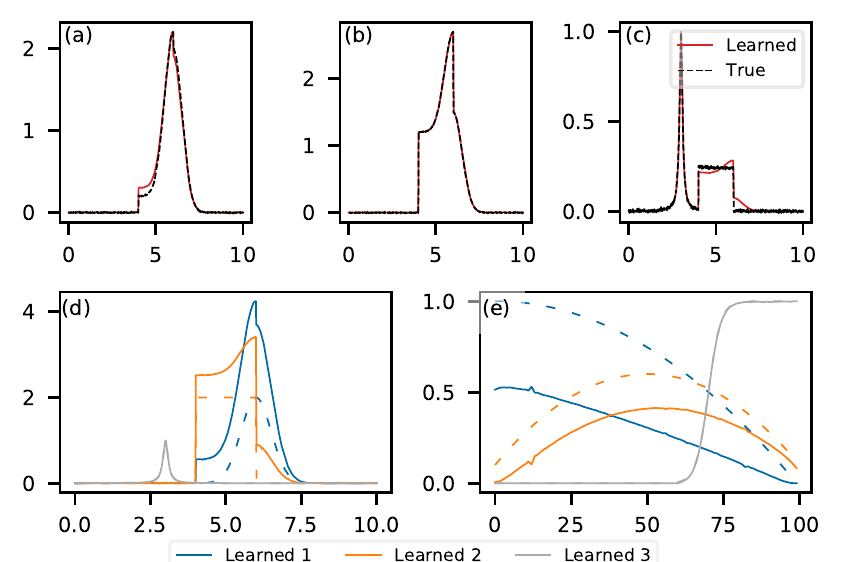}
    \caption{\label{fig:partial_2} The reconstruction of the (a) first, (b) median, and (c) last pattern of a dataset of mixed Gaussian, Lorenzian, and box functions using constrained NMF.
    (d) Prior knowledge of the Lorenzian function fixes one underlying component.
    (e) The learned weights are improved and create an better reconstruction of the dataset, than canonical NMF (MSE=3.838, R-factor=0.018).
    }
\end{figure}

\begin{figure}
    \includegraphics{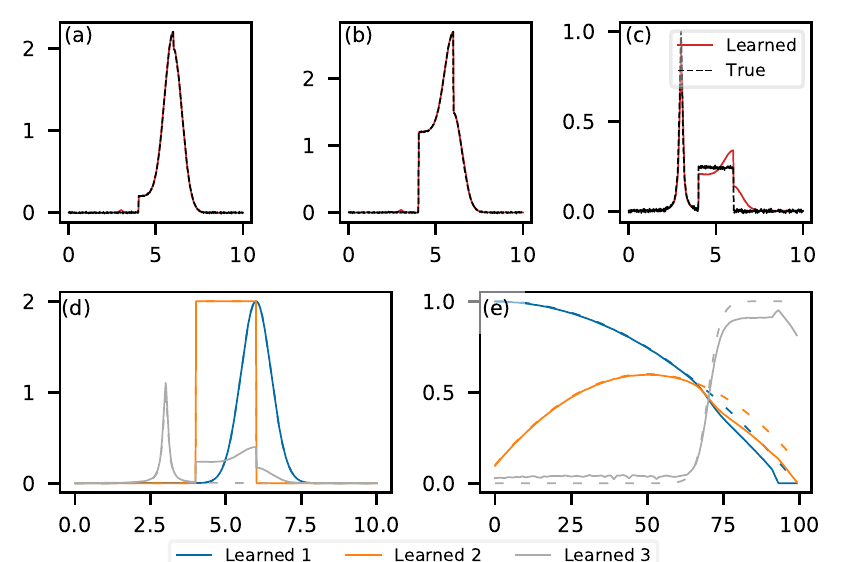}
    \caption{\label{fig:partial_01} The reconstruction of the (a) first, (b) median, and (c) last pattern of a dataset of mixed Gaussian, Lorenzian, and box functions using constrained NMF.
    (d) Prior knowledge of the Gaussian and box functions fixes two underlying components.
    (e) The learned weights are improved and create an better reconstruction of the dataset, than canonical NMF (MSE=3.788, R-factor=0.020).
    }
\end{figure}

\begin{figure}
    \includegraphics{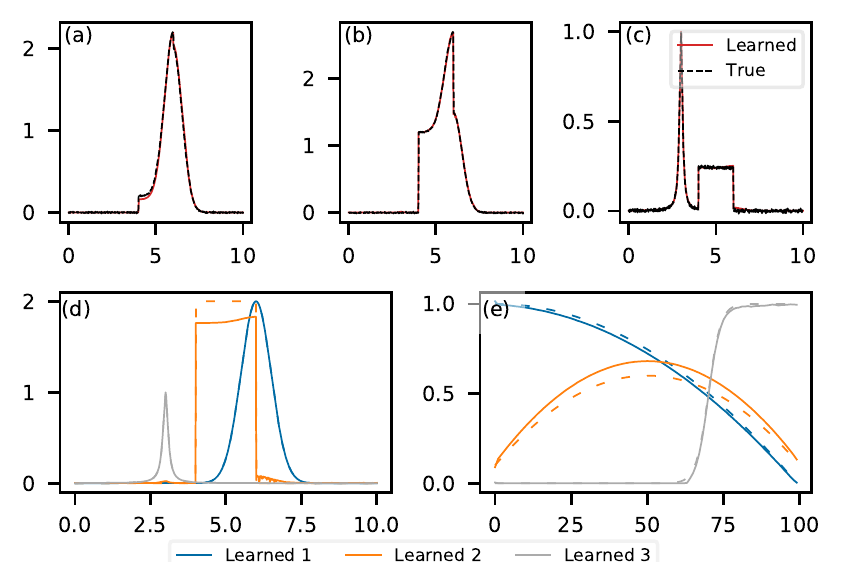}
    \caption{\label{fig:partial_02} The reconstruction of the (a) first, (b) median, and (c) last pattern of a dataset of mixed Gaussian, Lorenzian, and box functions using constrained NMF.
    (d) Prior knowledge of the Gaussian and Lorenzian functions fixes two underlying components.
    (e) The learned weights are improved and create an better reconstruction of the dataset, than canonical NMF (MSE=1.752, R-factor=0.015).
    }
\end{figure}

\begin{figure}
    \includegraphics{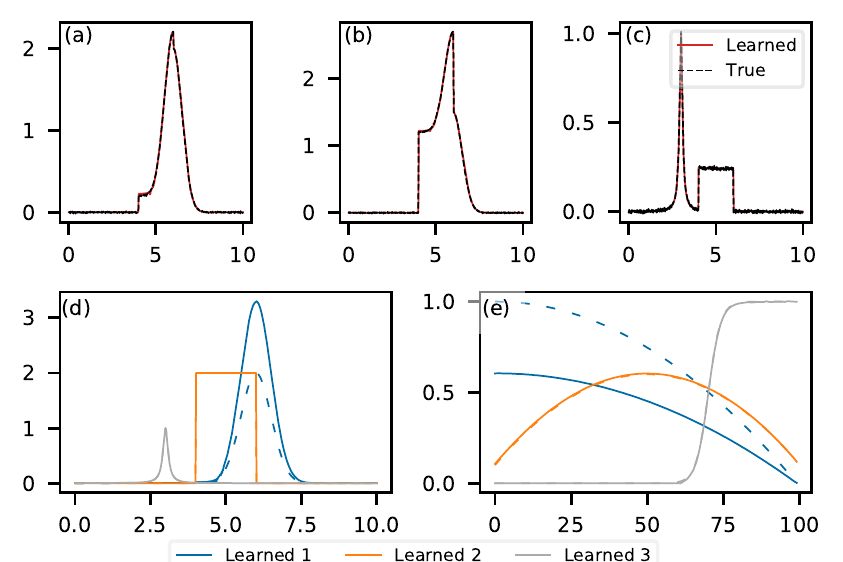}
    \caption{\label{fig:partial_12} The reconstruction of the (a) first, (b) median, and (c) last pattern of a dataset of mixed Gaussian, Lorenzian, and box functions using constrained NMF.
    (d) Prior knowledge of the box and Lorenzian functions fixes two underlying components.
    (e) The learned weights are improved and create an better reconstruction of the dataset, than canonical NMF (MSE=2.159, R-factor=0.015).
    }
\end{figure}

\section{Constraints enable physically meaningful \emph{in situ} analysis of melting  salts using the pair distribution function}

The results from the pair distribution function (PDF) analysis mirror that of the X-ray diffraction (XRD) results in the main text. The primary difference is the noise persistent in the resultant components, which is necessary to recreate the dataset. Since PDF is a calculated function from the integrated total scattering, it requires an additional step in the data stream to reduce the reciprocal space function into a real space PDF. This imperfect data reduction on-the-fly created a lack of uniformity at large radii for the amorphous regions. In this case, the function G(r) was used, pre-processed for NMF by min-max normalization. 

\begin{figure}
    \includegraphics{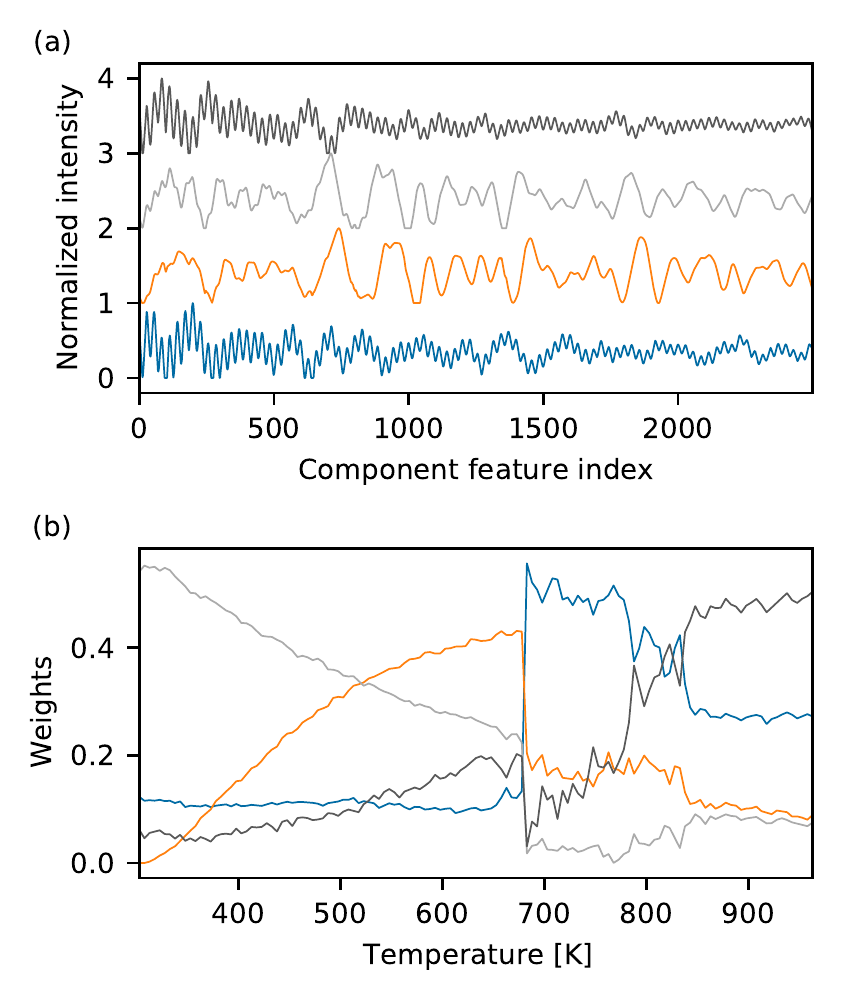}
    \caption{\label{fig:salt} Canonical NMF produces a confusing decomposition, offering limited clarity to the scientist.
    (a) The learned components are non-physical and cannot be interpreted as PDFs. 
    (b) The learned weights provide little information beyond an abrupt change around 400~\,K.
    }
\end{figure}

\begin{figure*}
    \includegraphics{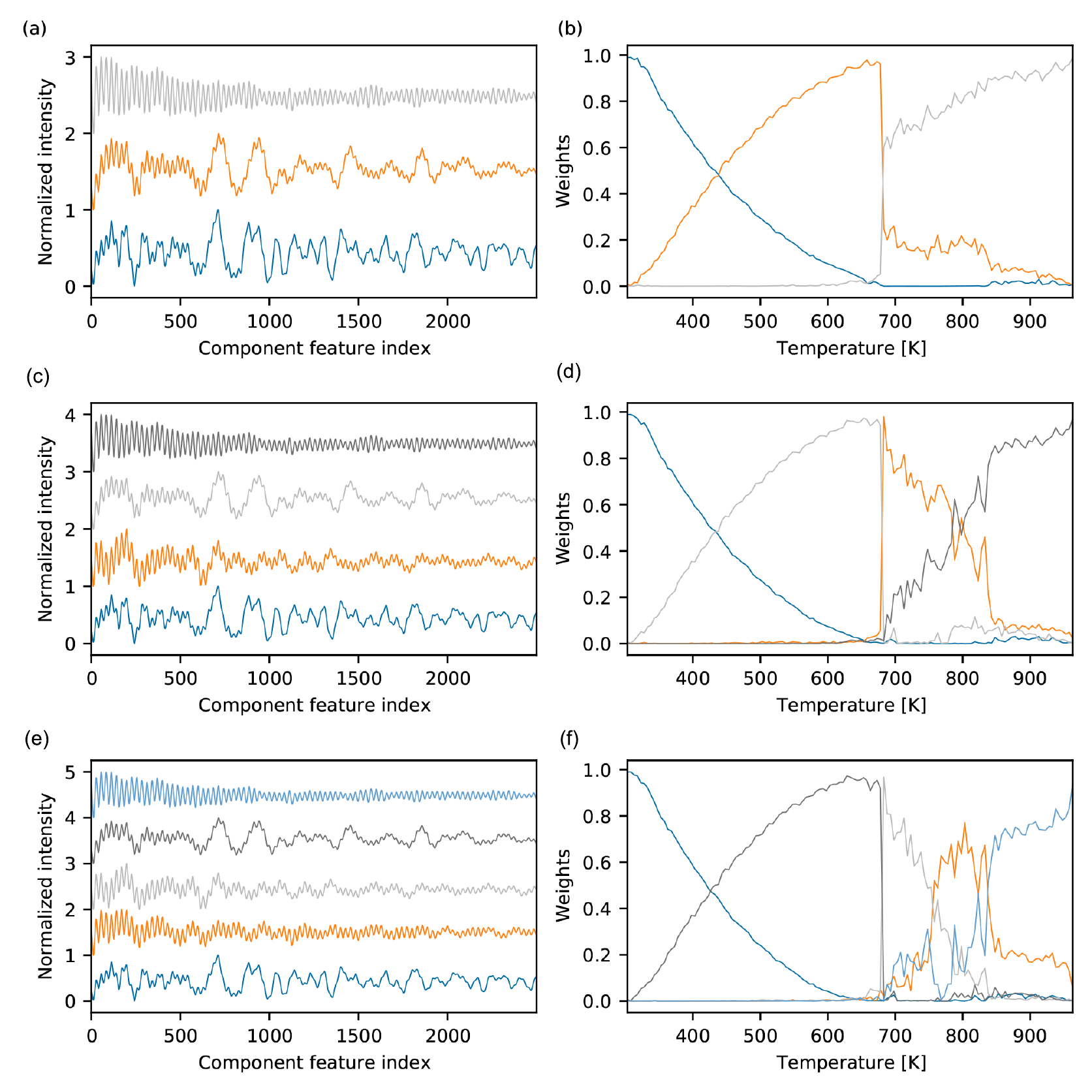}
    \caption{\label{fig:salt_summary} Automated constrained NMF was run using 3-6 components asynchronously on the molten salt dataset for PDFs.
    The learned components for the (a) 3, (c) 4, and (e) 6 component decompostions provide insight into thermal expansion, a phases present during the temperature ramp. 
    The learned weights for the  (b) 3, (d) 4, and (f) 6 component decompositions demonstrate the second-order phase transition, the characteristic response to thermal expansion, and persistence of a solid phase after 400~\,K.
    }
\end{figure*}